\documentclass[sn-mathphys-num 
              ]{sn-jnl}

\usepackage{graphicx}%
\usepackage{multirow}%
\usepackage{amsmath,amssymb,amsfonts}%
\usepackage{amsthm}%
\usepackage{mathrsfs}%
\usepackage[title]{appendix}%
\usepackage{xcolor}%
\usepackage{xspace}%

\usepackage{siunitx}%
\usepackage{lineno}%

\begin{document}
%
%
\title{Test of gravitational redshift by optical clocks at different latitudes}
\author[1]{\fnm{I.} \sur{Nosske}}
\equalcont{These authors contributed equally to this work.}
\author[1]{\fnm{C.} \sur{Vishwakarma}}
\equalcont{These authors contributed equally to this work.}
\author[1]{\fnm{T.} \sur{L{\"u}cke}}
\author[1]{\fnm{M.} \sur{Steinel}}
\author[1]{\fnm{M.} \sur{Filzinger}}
\author[1]{\fnm{J.} \sur{Kronj\"ager}}
\author[1]{\fnm{A.} \sur{Kuhl}}
\author[1]{\fnm{S.} \sur{Mukherjee}}
\author[1]{\fnm{E.} \sur{Benkler}}
\author[2]{\fnm{H.} \sur{Denker}}
\author[3,4]{\fnm{R.} \sur{Holzwarth}}
\author[3]{\fnm{Th.} \sur{Udem}}
\author[1]{\fnm{S.} \sur{D{\"o}rscher}}
\author[1]{\fnm{N.} \sur{Huntemann}}
\author*[1]{\fnm{C.} \sur{Lisdat}}\email{christian.lisdat@ptb.de}
\affil[1]{\orgname{Physikalisch-Technische Bundesanstalt}, \orgaddress{\street{Bundesallee 100}, \postcode{38116} \city{Braunschweig}, \country{Germany}}}
\affil[2]{Institut f\"ur Erdmessung, Leibniz Universit\"at Hannover, Schneiderberg 50, 30167 Hannover, Germany}
\affil[3]{Max-Planck-Institut f\"ur Quantenoptik, Hans-Kopfermann-Stra\ss e 1, 85748 Garching, Germany}
\affil[4]{Menlo Systems GmbH, Bunsenstra\ss e 5, 82152 Martinsried, Germany}
%
\newcommand{\AlexanderKuhl}{AK\xspace}
\newcommand{\ChetanVishwakarma}{CV\xspace}
\newcommand{\ChristianLisdat}{CL\xspace}
\newcommand{\ErikBenkler}{EB\xspace}
\newcommand{\HeinerDenker}{HD\xspace}
\newcommand{\IngoNosske}{IN\xspace}
\newcommand{\JochenKronjaeger}{JK\xspace}
\newcommand{\MartinSteinel}{MS\xspace}
\newcommand{\MelinaFilzinger}{MF\xspace}
\newcommand{\NilsHuntemann}{NH\xspace}
\newcommand{\RonaldHolzwarth}{RH\xspace}
\newcommand{\ShamboMukherjee}{SM\xspace}
\newcommand{\SoerenDoerscher}{SD\xspace}
\newcommand{\ThomasUdem}{TU\xspace}
\newcommand{\TimLuecke}{TL\xspace}
\abstract{\unboldmath
    General relativity predicts that frequency comparisons between clocks are affected by both gravitational and kinematic time dilation.
    While the kinematic contribution has been tested with high accuracy, experimental tests of the gravitational redshift remain comparatively less stringent.
    Here, we measure the frequency shift between two Earth-fixed optical atomic clocks at different potentials and compare it to the redshift predicted by general relativity using independent geodetic measurements.
    The observed shift agrees with the prediction, testing the gravitational redshift at the \mbox{$4.6 \times 10^{-5}$} uncertainty level, improving ground-based tests by a factor two.
    Previous ground-based tests separated the clocks mainly vertically.
    For clocks at different latitudes, their height difference is governed not only by the gravitational potentials at their locations but also by the centrifugal potentials caused by Earth's rotation.
    As both contributions can partially cancel, the gravitational contribution can be considerably larger than their height difference suggests.
    In this work, a separation of the clocks by $4^\circ$ in latitude enhances their sensitivity to a gravitational redshift violation threefold, to the equivalent of an approximately 1200~m difference in height for small latitudinal separation.
    Our results establish Earth’s rotation as a so far unexploited resource for advancing ground-based tests of the gravitational redshift towards the sensitivity of previous space-based experiments.
}

\maketitle
%
%
%
%
General relativity (GR) \cite{ein16} describes gravitation as a manifestation of spacetime curvature and has withstood a wide range of stringent experimental tests for several decades \cite{wil14}. 
Despite its remarkable success, the theory is believed to be incomplete because it is incompatible with quantum mechanics and because dark matter \cite{ber18e} and dark energy \cite{bra17a} need to be added to explain observed astronomical and cosmological patterns of the universe \cite{bah99}.
Therefore, it remains necessary to test GR and the underlying Einstein equivalence principle \cite{dam99} through measurements of physical observables such as the gravitational redshift (GRS).

The most stringent tests of Local Position Invariance by measurements of the GRS to date have been obtained by comparing atomic clocks \cite{del18, her18b, tak20}.
Conceptually, they measure the frequency shift $\Delta\nu$ between two clocks separated by a Newtonian gravitational potential difference $\Delta U$, which is given by
\begin{equation}
    \frac{\Delta \nu}{\nu} = (1 + \alpha) \frac{\Delta U}{c^2}
    \label{grs:very-simple}
\end{equation}
to first order in $\Delta U$, where $\nu$ is the clock frequency, $c$ is the speed of light, and $\alpha$ parametrizes a possible deviation from GR.
Space-based measurements using microwave clocks aboard Galileo satellites in eccentric Earth orbits have constrained $\alpha$ with an uncertainty \mbox{$u(\alpha) = 2.5 \times 10^{-5}$} \cite{del18}.
Ground-based experiments benefit from the higher accuracies of state-of-the-art optical clocks,
but are restricted to much smaller potential differences between them.
Previously, the tightest constraint in a ground-based experiment had been achieved in the Tokyo Skytree experiment with \mbox{$u(\alpha) = 9.1 \times 10^{-5}$} \cite{tak20}.

\begin{figure*}[t]
    \centering
    \makebox[\textwidth][c]{
        \includegraphics[width=\textwidth]{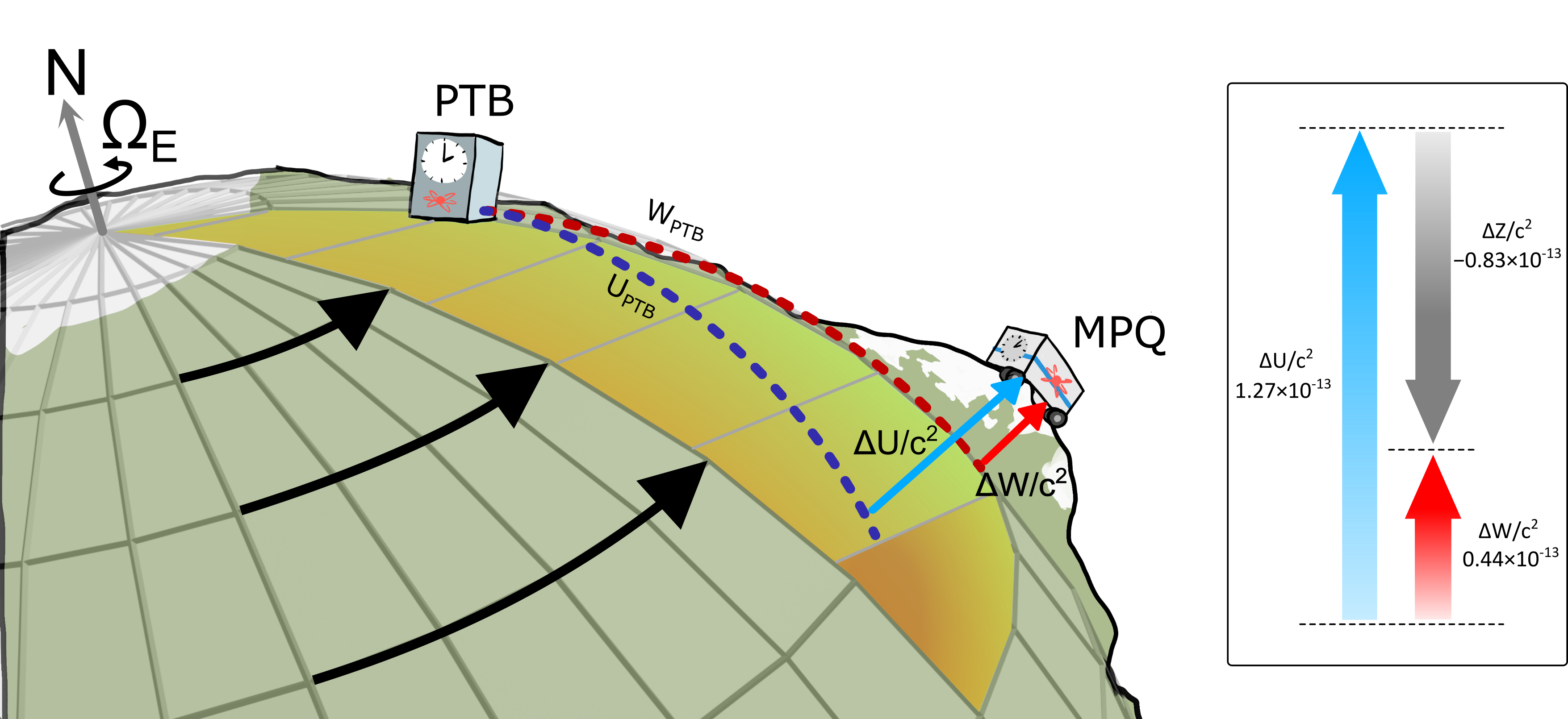}
    }
    \caption{
        \textbf{Schematic of the gravitational redshift (GRS) test with enhanced sensitivity.}
        We measure the relativistic redshift (RRS) of a clock's frequency between two stations at different heights and latitudes (labelled MPQ and PTB, see main text for details).
        The frequency shift is measured by combining a remote measurement of the frequency ratio between a transportable clock at MPQ and a reference clock at PTB with their co-located frequency ratio measured at PTB.
        The result is compared to the prediction of general relativity in order to test for a GRS violation.
        The red (blue) dashed line represents the gravity (gravitational) equipotential surface, on which PTB is situated.
        Since MPQ is located higher and at a lower latitude than PTB, the GRS ($\Delta U / c^2$, blue arrow) is larger than the net RRS ($\Delta W / c^2 = (\Delta U + \Delta Z) / c^2$, red arrow) owing to partial cancellation by the kinematic term ($\Delta Z / c^2$, grey arrow), as shown in the inset.
        This enhances the GRS test's sensitivity by a factor of about three compared to a local GRS test with the same height difference.
    }
    \label{fig:schematic}
\end{figure*}

\begin{figure}[t]
    \centering
    \includegraphics[width=0.5\linewidth]{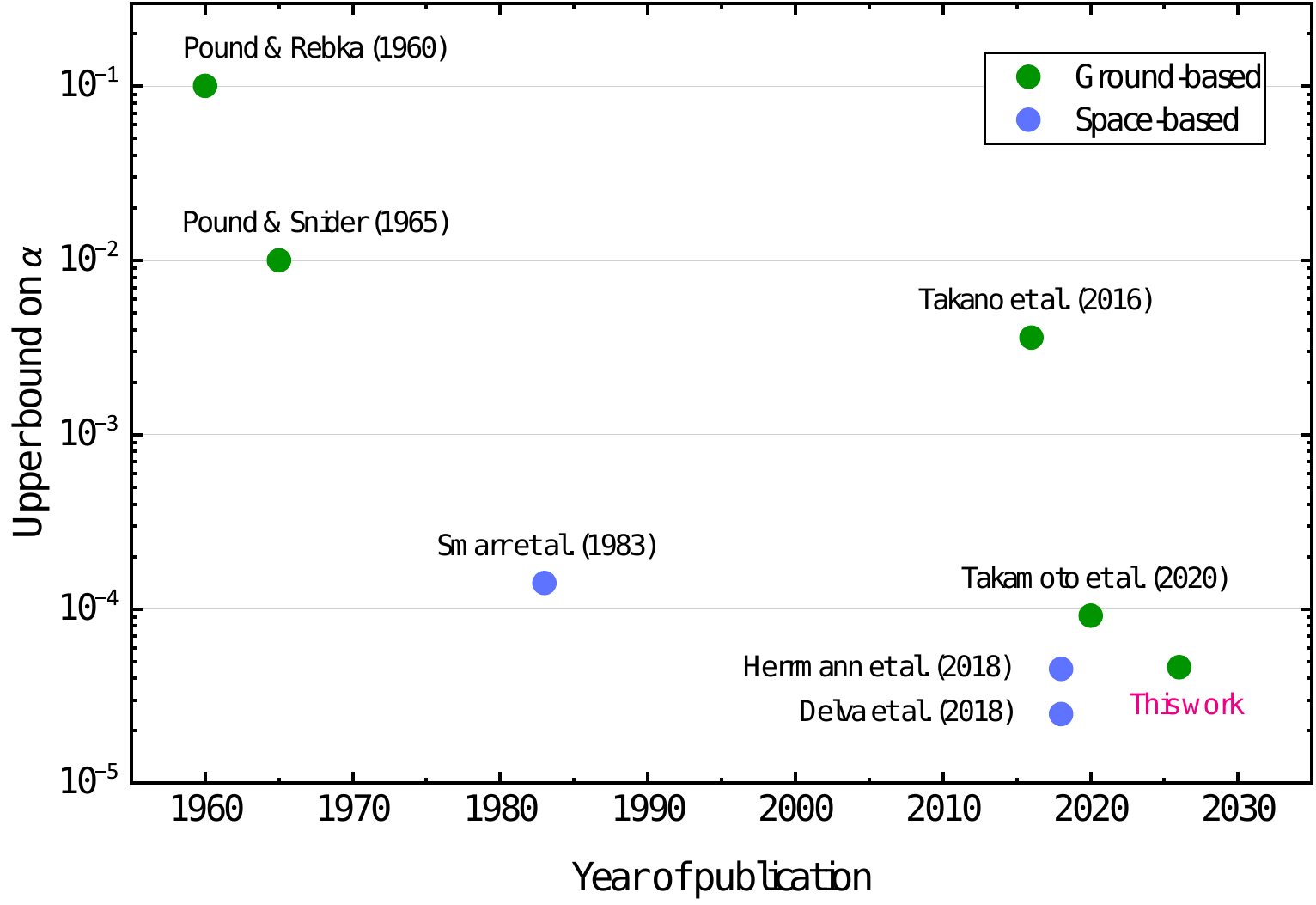}
    \caption{
        \textbf{Results of GRS tests.}
        Upper bounds on the violation parameter $\alpha$ reported for ground-based \cite{pou60, pou65, tak16, tak20} and space-based \cite{sma83, del18, her18b} tests by time of publication.
        The points represent the reported value for Ref.~\cite{her18b}, as it exceeds the uncertainty, and the reported uncertainties otherwise.
    } 
    \label{fig:AlphaOverViewPlot}
\end{figure}

Here, we demonstrate how Earth's rotation can be leveraged to improve the sensitivity of ground-based GRS tests and report a test with \mbox{$u(\alpha) = 4.6 \times 10^{-5}$} by comparing optical clocks between the Max Planck Institute for Quantum Optics (MPQ) in Garching, Germany and the Physikalisch-Technische Bundesanstalt (PTB) in Braunschweig, Germany via an interferometric fibre link, as shown in Fig.~\ref{fig:schematic}.
This result improves ground-based test of the GRS \cite{pou60, pou65, tak16, gro18a, tak20} by a factor two and is nearly as stringent as space-based measurements \cite{sma83, del18, her18b} (Fig.~\ref{fig:AlphaOverViewPlot}).

\section*{Improving the sensitivity of ground-based GRS tests}
The relativistic redshift (RRS) between Earth-fixed clocks is not given solely by the gravitational term (i.e., the GRS), but also contains a kinematic term that stems from Earth's rotation.
In a geocentric, non-rotating reference frame, the latter is a relativistic Doppler shift and is well described by a centrifugal potential
\begin{equation}
    Z = -\frac{1}{2} v^2 = -\frac{1}{2}\Omega_E^2 p^2
    \label{eq:centrifugal-potential}
\end{equation}
to first order in $v^2$, where $v$ is the velocity of the Earth-fixed clock resulting from the angular velocity $\Omega_E$ of Earth's rotation at a distance $p$ from the axis of rotation.
The gravitational potential $U$ and the centrifugal potential $Z$ are often combined into the gravity potential
\begin{equation}
    W = U + Z \mbox{.}
    \label{eq:gravity-potential}
\end{equation}
Notably, heights as commonly used and the gravity acceleration are connected to this combined potential $W$ instead of the gravitational potential $U$.
Similar to space-based experiments \cite{sma83, del18, her18b, mar23}, the frequency shift is then given by
\begin{eqnarray}
    \frac{{\Delta \nu}}{\nu} & = & (1 + \alpha) \frac{\Delta U}{c^2}
                                 + (1 + \alpha_\mathrm{SR}) \frac{\Delta Z}{c^2} \nonumber\\
                             & = & \frac{\Delta W}{c^2}
                                 + \alpha \frac{\Delta U}{c^2}
                                 + \alpha_\mathrm{SR} \frac{\Delta Z}{c^2}\mbox{,}
    \label{eq:RelTest}
\end{eqnarray}
where $\alpha_\mathrm{SR}$ parametrizes a violation of Lorentz Invariance and has been constrained to $\left|\alpha_\mathrm{SR}/2\right| \le 1.1 \times 10^{-8}$ by comparisons of optical atomic clocks \cite{del17}.

Previous ground-based GRS tests \cite{tak16, tak20} have been performed between optical clocks separated by no more than about 15~km horizontally and 500~m vertically.
The differences between the RRS and GRS could be neglected (${\Delta W} \approx {\Delta U}$) as the changes in centrifugal potentials were small across such separations.
For large latitudinal separations, however, this approximation is usually no longer suitable because the centrifugal potential depends on the latitude $\varphi$, reaching $(\partial Z / \partial\varphi) / g \approx 0.2~\mathrm{km}/{}^\circ$ at $\varphi \approx 45{}^\circ$.
The centrifugal potential difference $\Delta Z$ can easily be comparable to the gravitational potential difference $\Delta U$ or even larger than it in magnitude.

If one clock is located both higher and equatorward with respect to the other, the GRS ${\Delta U} / c^2$ exceeds the RRS ${\Delta W} / c^2$, since $\Delta W > 0$, $\Delta Z < 0$ and thus $\Delta U = {\Delta W} - {\Delta Z} > {\Delta W}$.
The sensitivity of a GRS test in this geographical configuration is then enhanced by the factor ${\Delta U}/{\Delta W}$ compared to a test in close proximity (i.e., $\Delta Z \approx 0$) with the same gravity potential difference.
This is because the uncertainty contributions from a possible violation of Lorentz Invariance and from the centrifugal potential difference $\Delta Z$ to the test's total uncertainty $u(\alpha)$ are negligible (see Methods for details).

\section*{Geodetic Measurements}
In this work, we investigate the RRS between optical atomic clocks located at PTB (\mbox{$\varphi \approx 52.3^\circ$}) and MPQ (\mbox{$\varphi \approx 48.3^\circ$}).
In order to establish the GR prediction, we first determined the gravity potentials at the clocks' locations by independent geodetic measurements, connecting the clocks to height reference markers at each station through geometric levelling and the two distant markers through the ``GNSS/geoid approach'' \cite{den17, rie20}.
The sensitivity of the geodetic measurements to possible violations of the GRS is negligible.
Further details are given in the Methods.
We find \mbox{$\Delta W / c^2= 436\,607(35) \times 10^{-19}$}, i.e., the clock is about 400~m higher at MPQ than at PTB.
Here, the uncertainty $u(\Delta W)$ is limited by the geoid model.

We further determined the  relativistic Doppler shifts for the clocks' coordinates using the known rotation and shape parameters of Earth (see Methods), obtaining \mbox{$\Delta Z / c^2 = -834\,454(15) \times 10^{-19}$}.
The uncertainty $u(\Delta Z)$ is dominated by the latitudinal position uncertainties of the clocks, but only has minor influence on the result of the GRS test (Methods).
Using Eq.~(\ref{eq:gravity-potential}), the gravitational potential difference is thus given by $\Delta U / c^2 = 1\,271\,061(38) \times 10^{-19}$.
Together with the above-mentioned constraint on $\alpha_\mathrm{SR}$ from previous measurements \cite{del17}, this determines all parameters on the right-hand side of Eq.~(\ref{eq:RelTest}) except the violation parameter $\alpha$ and thus allows us to constrain the latter by a measurement of the frequency shift $\Delta\nu/\nu$ between the clocks.
Here, the separation of the clocks by about $4^\circ$ in latitude enhances the sensitivity of the GRS test nearly threefold ($\Delta U / \Delta W \approx 2.9$) compared to a local GRS test with the same height difference of about 400~m or a test of the total RRS.

\section*{Optical clock comparisons}
We measure the frequency shift $\Delta\nu/\nu$ between MPQ and PTB experienced by a transportable \textsuperscript{87}Sr optical lattice clock.
A stationary \textsuperscript{171}Yb\textsuperscript{+} single-ion clock serves as frequency reference.
Clock comparisons of our transportable clock Sr4 \cite{nos25} and our laboratory single-ion clock Yb1E3 \cite{hun16, san19} at PTB over the course of nearly two years have been reported previously \cite{vis26}.
Both clocks have been evaluated to fractional frequency uncertainties below $3 \times 10^{-18}$ \cite{san19, nos25}.
These measurements have established the co-located frequency ratio at PTB at the level of $4.3 \times 10^{-18}$ relative uncertainty.
For this work, we performed an additional, remote clock comparison between these clocks in a six-week measurement campaign, during which Sr4 was deployed at MPQ.
The two clocks were connected to each other by a fibre link with a length of about 940~km, which enabled transmission of phase-stable laser light at a frequency of about $194~\mathrm{THz}$ between the stations.
On each end, an optical frequency comb measured the frequency ratio of the light from the fibre link and the local clock.
Co-located comparisons were carried out before and after the campaign  \cite{vis26} to ensure that, apart from the RRS, the frequency ratio is reproduced within the estimated uncertainty.
Further details on the remote clock comparisons and the influence of local RRS corrections are given in the Methods.

\begin{figure*}[t]
    \centering
    \includegraphics[width=0.49\linewidth]{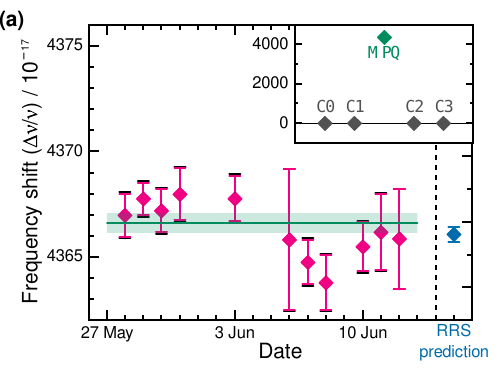}
    \includegraphics[width=0.49\linewidth]{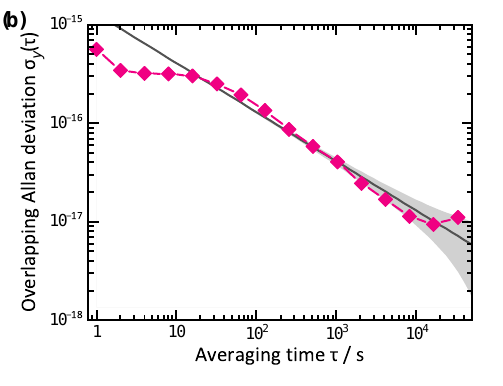}
    \caption{
        \textbf{Results of the remote clock comparison between the transportable $^{87}$Sr lattice clock at MPQ and the $^{171}$Yb$^{+}$ ion clock at PTB.}
        \textbf{(a)}
        \textbf{Fractional frequency shift of the transportable clock:}
        The panels show mean fractional frequency shifts of the transportable clock with respect to the mean co-located value determined in Ref.~\cite{vis26}.
        In the main panel, magenta diamonds and magenta (black) error bars represent the daily means during the remote campaign at MPQ in 2024 and their statistical (total) standard uncertainties, respectively.
        The blue diamond shows the GR prediction $\Delta W / c^2$ based on independent geodetic measurements.
        The solid green line and shaded band represent the arithmetic mean of all valid one-second remote data points and its total standard uncertainty, respectively.
        The inset shows this mean value (MPQ, green) and those of the individual co-located campaigns (C0--C3, gray) reported in Ref.~\cite{vis26}.
        \textbf{(b)}
        \textbf{Fractional instability of the frequency ratio:}
        Magenta diamonds show the overlapping Allan deviation of the concatenated valid one-second data points.
        The solid black line and shaded band represent a white-frequency-noise instability of $\num{1.3e-15} / \sqrt{\tau / \si{\second}}$ and the \qty{68.3}{\percent} central probability interval for a data set with this white frequency noise amplitude and the same length as the measurement, respectively \cite{ril08, gre03b}.
    }
    \label{fig:Daily remote ratio}
\end{figure*}

\section*{Results}
The remote comparison yielded about $1.6 \times 10^5~\mathrm{s}$ of valid measurement data across eleven days.
Figure~\ref{fig:Daily remote ratio}a shows the observed daily means of the frequency shift, where the statistical uncertainty of each dataset is determined by extrapolating the Allan deviation in the white-frequency-noise regime to the full dataset.
The reduced chi-squared value of the daily means ($\chi_\mathrm{red}^2 = 1.45$) does not indicate the presence of excess scatter beyond the estimated statistical uncertainties \cite{moh00}.

Across the entire campaign, we measure a mean frequency shift $\Delta \nu / \nu = 436\,661(47) \times 10^{-19}$.
Here, the largest uncertainty contribution stems from the fractional statistical uncertainty of the remote measurement at $3.2 \times 10^{-18}$, which results from the instability in the white-frequency-noise regime, $\sigma_y(\tau) \approx 1.3 \times 10^{-15}/ \sqrt{\tau/\mathrm{s}}$, where $\tau$ is the averaging time (see Fig.~\ref{fig:Daily remote ratio}b).
The fractional statistical uncertainty of the co-located measurement is $2.1 \times 10^{-18}$.
The fractional systematic uncertainty contributions from the transportable and reference clock are $2.6 \times 10^{-18}$ and $0.7 \times 10^{-18}$, respectively, taking into account the correlations between the local and remote measurements (see Methods).
This result improves upon an earlier measurement between the same stations \cite{gro24} by more than a factor of six.

Solving Eq.~(\ref{eq:RelTest}) for the violation parameter $\alpha$ and inserting the values given above for the other quantities, we then find
\begin{equation*}
    \alpha = 4.3(4.6) \times 10^{-5}
\end{equation*}
for our measurements.
Since this value is compatible with zero within its uncertainty, we find no deviation from the GRS predicted by GR.
Our result improves upon the previous best uncertainty of $9.1 \times 10^{-5}$ reported for a ground-based GRS test \cite{tak20} by about a factor of two.
It is also within a factor of two of the best uncertainty of $2.5 \times 10^{-5}$ reported for any GRS test \cite{del18} (see Fig.~\ref{fig:AlphaOverViewPlot}).

The uncertainty of the measured violation parameter $\alpha$ is dominated by the contribution from the clock comparisons ($u(\alpha_{\Delta\nu / \nu}) = 3.7 \times 10^{-5}$).
Contributions from a possible violation of Lorentz Invariance ($u(\alpha_{\alpha_\mathrm{SR}}) = 1.4 \times 10^{-8}$) and the determination of centrifugal potential difference ($u(\alpha_{\Delta Z / c^2}) = 5.2 \times 10^{-10}$) are negligible.
The contribution from the gravity potentials obtained using the GNSS/geoid approach ($u(\alpha_{\Delta W/c^2}) = 2.7 \times 10^{-5}$), however, is comparable to the contribution from optical clock comparisons and thus limits further improvement even as optical clock performance advances.

\section*{Extension of the measurement scheme}
\begin{figure}[t]
    \centering
    \includegraphics[width=0.5\linewidth]{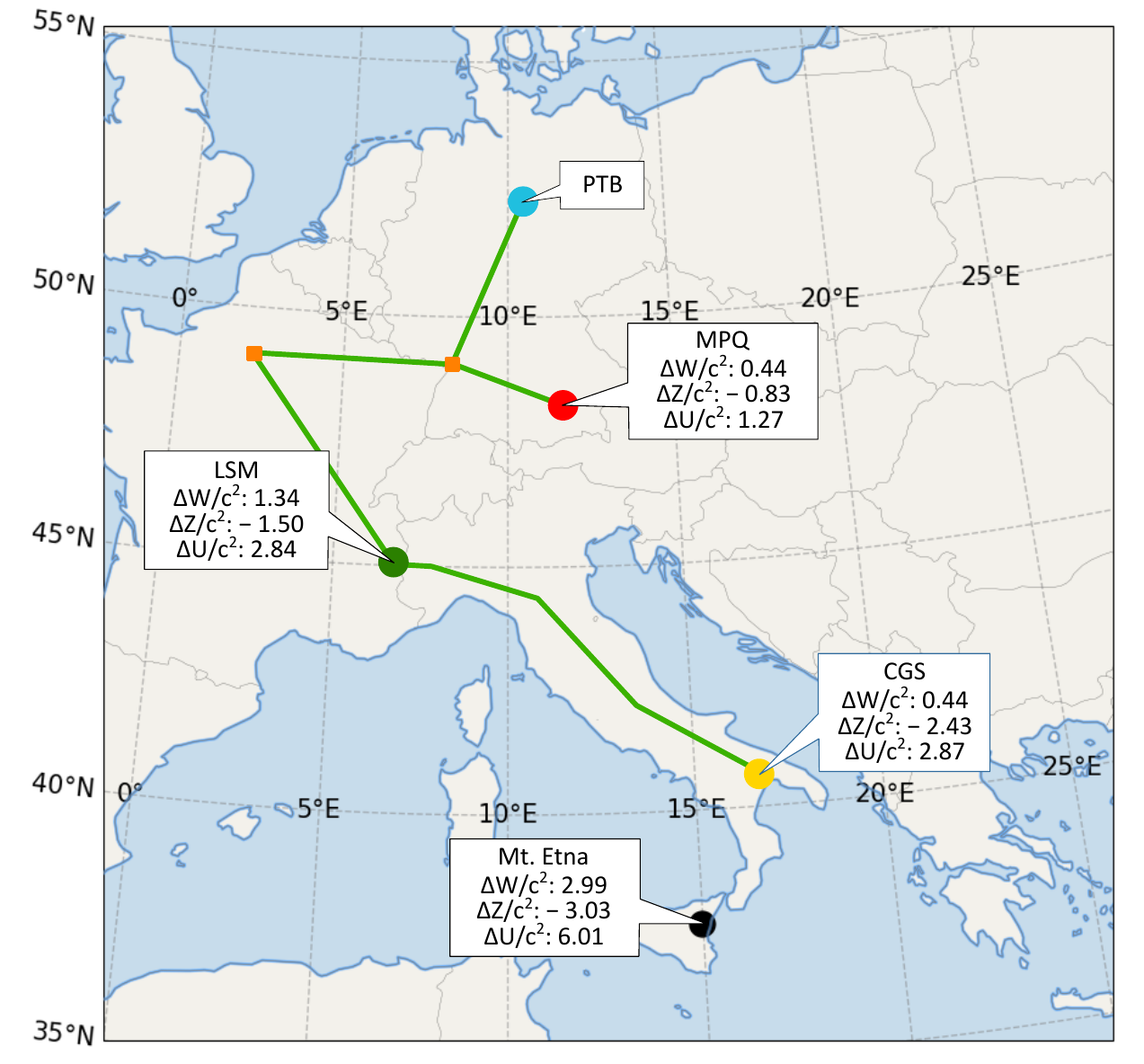}
    \caption{
        \textbf{Locations of interest for gravitational redshift tests by optical clocks in Europe.}
        Total ($\Delta W/c^2$), relativistic Doppler ($\Delta Z/c^2$) and gravitational (\mbox{$\Delta U/c^2$}) redshifts of selected sites in Europe relative to PTB (values given in units of $10^{-13}$).
        Except Mount Etna, all locations are connected to PTB via fibre links \cite{lis16, cli22b, can24}.
    } 
    \label{fig:InstituteRedshifts}
\end{figure}

Nevertheless, continent-spanning fibre links like the European optical network \cite{lis16, cli22b, can24} can enable improved sensitivity of ground-based GRS tests, by increasing the leverage.
Figure~\ref{fig:InstituteRedshifts} shows the potential differences for a number of stations across Europe with respect to PTB, which is a favourable reference station due to its high latitude and low height.
Stations that are located higher, equatorward, or both offer comparatively high gravitational potential differences $\Delta U$ as leverage.
For instance, the leverage for both the Laboratoire Souterrain de Modane (LSM) in the French Alps and the Giuseppe Colombo Centre for Space Geodesy (CGS) in southern Italy is about twice as large as for MPQ.
Both stations have already also been connected to the fibre link network \cite{lis16, cli22b, can24}.
Further improvement by more than another factor of two is possible, e.g., if the network is extended to Etna Observatory in Sicily.
The uncertainties of the gravity potential differences determined by the GNSS/geoid approach are largely independent of distance for non-local clock comparisons.

Outside Europe, there are currently no fibre links that connect stations with comparable differences in latitude.
The closest example is a link in Japan connecting RIKEN in the Greater Tokyo Area to the National Astronomical Observatory of Japan (NAOJ) in Mizusawa \cite{ish25a}.
These two stations are separated by $3.4^\circ$ in latitude and differ by \mbox{$\Delta U / c^2 \approx 0.6 \times 10^{-13}$}.

\section*{Outlook}
While space-based tests offer substantially increased sensitivities \cite{lam02, cac09, sav19a} due to their larger accessible modulation in gravitational potential, ground-based GRS tests will remain highly relevant as they can be performed more frequently and sustainably and allow the use of the latest available technology.
A reduction in the clock measurement uncertainty by another order of magnitude appears feasible in the medium term \cite{mcg18, arn26a}, while measurement sites on high mountains are becoming accessible via free-space frequency transfer techniques \cite{she22, goz22}.
The sensitivity of a GRS test at a high mountain could be further enhanced by additionally exploiting the Earth’s rotation along a long latitudinal baseline, as demonstrated in this article.
Ground-based optical clock comparisons can thus serve as sensitive probes to investigate fundamental aspects of gravitation and to search for deviations from GR (e.g., through increasingly stringent tests of the gravitational redshift and the Einstein equivalence principle), and provide a robust basis for high-precision chronometric levelling  \cite{meh18, icon24}.

\section*{Methods}

\subsection*{Expressing frequency shifts through frequency ratios}
While the RRS experienced by the transportable clock Sr4 is expressed as a relative frequency shift $\Delta\nu/\nu$ throughout this work, it is determined from separate co-located and remote measurements of the optical frequency ratio with the reference clock Yb1E3 at PTB, which uses an atomic transition at a different frequency.
The frequency shift is related to the co-located frequency ratio $R = \nu(\mathrm{{}^{87}Sr}) / \nu(\mathrm{^{171}Yb^+, E3})$ and the remote frequency ratio $R^\prime = (1 + \Delta\nu/\nu) R$, which includes the RRS observed in those measurements, by $\Delta\nu/\nu = (R^\prime - R) / R \equiv \Delta R / R$.

\subsection*{Determination of the gravity potential difference}
The relevant gravity potentials are those of the transportable clock Sr4 at MPQ and the reference clock Yb1E3 at PTB.
We neglect the effect of a potential GRS violation through the local RRS between the clocks at PTB \cite{vis26}, which is corrected using the GR prediction in the measurements at PTB.
The gravity potential difference $\Delta W$ between the transportable clock at MPQ and the reference clock at PTB is then given by the sum of the following contributions, which are expressed in units of $c^2$ for simplicity.

\paragraph{Mean potential difference between markers at MPQ and PTB}
The mean gravity potentials of the height reference markers MPQ-2011 \cite{den19} and PTB-KB02 \cite{rie20} have been determined previously via the GNSS/geoid approach, which combines ellipsoidal heights from global navigation satellite systems (GNSS) with a geoid model, with additional terrestrial gravity measurements for refinement \cite{den17}.
Here, the (quasi)geoid model EGG2015 and the coordinate reference frame ITRF2008 with the epoch \mbox{2005.0} are used.

These determinations report zero-tide geopotential numbers \cite{mae21} for the reference markers, i.e., they exclude the time-averaged tidal potential of other celestial bodies.
We therefore correct their difference by \mbox{$-22 \times 10^{-19}$} with negligible uncertainty to transform it to the mean-tide difference \cite{ihd08}.

Overall, we find a contribution of $436\,447(33) \times 10^{-19}$, which is the dominant contribution to the value and uncertainty of $\Delta W$.

\paragraph{Local potential differences}
The potential differences between each clock and its co-located height reference marker are determined by combining the height difference with the mean gravity acceleration at each station.
At PTB, the reference clock is 1.001(3)~m above KB02 as determined using a combination of spirit levelling and electronic laser levelling, resulting in a contribution of $-1093(3) \times 10^{-19}$ to the potential difference.
At MPQ, we find that the transportable clock is 1.145(9)~m above the marker 2011 using a self-levelling line laser instrument.
Here, the transportable clock's trailer is located about $20~\mathrm{m}$ from the reference marker. 
The contribution to the potential difference is $1250(10) \times 10^{-19}$.

\paragraph{Dynamic correction}
Sampling of the time-varying tidal components of the gravity potential \cite{voi16} causes an additional contribution.
We account for the time-varying components using the model described in Ref.\ \cite{voi16}, where the dynamic effects are mainly the tidal potential and solid Earth tides, and to a lesser extent ocean tidal loading, the pole tide and length-of-day tides.
Overall, we find a mean contribution of \mbox{$+4 \times 10^{-19}$} over our measurement, with negligible uncertainty.

\subsection*{Determination of the relativistic Doppler shift difference}
The centrifugal potential of an Earth-fixed point due to the planet's rotation is given by Eq.~(\ref{eq:centrifugal-potential}), where $\Omega_\mathrm{E} = 7.292 \, 115 \times 10^{-5} ~ \mathrm{rad/s}$.
The distance of the point from the axis of rotation is given by \cite{tor75}
\begin{equation}
    p = (N + h) \cos\varphi~\mbox{,}
\end{equation}
where $h$ and $\varphi$ are the point's ellipsoidal height and latitude, respectively, with respect to the reference ellipsoid of the Geodetic Reference System 1980 (GRS80).
The radius of curvature in the prime vertical, $N$, is given by
\begin{equation}
    N = a {\left( 1 - e^2 \sin^2\varphi \right)^{-1/2}}\mbox{,}
\end{equation}
with the semi-major axis $a = 6\,378\,137~\mathrm{m}$ and the squared eccentricity $e^2 = 2f - f^2$, where $f = 1/298.257\,222\,101$ is the flattening.
Here, the values of $\Omega_\mathrm{E}$, $a$, and $f$ are conventional constants from the GRS80 reference ellipsoid \cite{mor80}.

For the reference clock, we obtain $\varphi_\mathrm{PTB} = 52^\circ$ $17'$ $46.30(19)''$ and $h_\mathrm{PTB} = 119.71(1)~\mathrm{m}$, resulting in $p_\mathrm{PTB} = 3\,909 \, 010.4(4.7)~\mathrm{m}$ and $Z_\mathrm{PTB} / c^2 = -4\,520\,321(11) \times 10^{-19}$.
For the transportable clock, $\varphi_\mathrm{MPQ} = 48{}^\circ$ $15'$ $35.80(19)''$, $h_\mathrm{MPQ} = 522.31(1)~\mathrm{m}$, $p_\mathrm{MPQ} = 4\,254 \, 541.9(4.4)~\mathrm{m}$ and $Z_\mathrm{MPQ} / c^2 = -5\,354\,775(11) \times 10^{-19}$.
Here, the uncertainty in latitude is the dominant contribution to the uncertainty of the centrifugal potential;
it consists of contributions from the position of the atoms or ion relative to the respective height marker \cite{den19, rie20} ($0.16''$, corresponding to $5.0~\mathrm{m}$) and from the available resolution of the height marker's ellipsoidal latitude ($0.1''$).
The uncertainty contributions from the ellipsoidal heights and the angular velocity of the Earth (\mbox{$u(\Omega_\mathrm{E}) = 2 \times 10^{-12} ~ \mathrm{rad/s}$}) are negligible.
Here, the latter accounts for variations of the Earth's rotation rate, calculated from length-of-day data of the IERS Earth orientation parameters C04 \cite{IERS_EOP20C04_1962now} during the measurement period.
These conservative estimates of the centrifugal potentials' uncertainties are sufficient because the resulting uncertainty in $\Delta Z$ is negligible for the GRS test (see main text and below).
We note that achieving lower uncertainties for the latitudinal positions of the clocks would  have been feasible, but was not necessary.

\subsection*{Optical clock comparison}
\paragraph{Transportable optical clock}
The transportable clock Sr4 is an optical lattice clock that interrogates the 429~THz clock transition (\mbox{$^1\mathrm{S}_0$--${^3\mathrm{P}_0}$}) of ${}^{87}$Sr atoms within a blackbody radiation shield, at a temperature close to $-50 \, ^{\circ}\mathrm{C}$ for the reported measurement.
It has been described in detail in Ref.\ \cite{nos25}.
At MPQ, its systematic frequency uncertainty was evaluated to $u(\nu_\mathrm{Sr4}) / \nu_\mathrm{Sr4} = 3.3\times10^{-18}$, which is larger than the previously reported value of \num{2.1e-18} \cite{nos25, vis26} due to increased contributions from the density and lattice light shifts during this remote campaign.
The clock was housed in an air-conditioned car trailer, except for the clock laser \cite{her22}, which was situated in a laboratory at MPQ.
The clock laser operated on the first subharmonic of the clock transition.
We interrogated the clock transition using Rabi spectroscopy with a $\pi$-pulse duration of 500~ms.
This results in an interleaved self-comparison instability of \mbox{$9 \times 10^{-16}/ \sqrt{\tau/\mathrm{s}}$} over the data set recorded at MPQ, while the transportable clock's contribution to the instability of the comparison is expected to be lower than that.

\paragraph{Reference optical clock}
The reference clock Yb1E3 is a laboratory clock that interrogates the 642~THz electric-octupole (E3) transition (\mbox{${}^2\mathrm{S}_{1/2}(F=0)$--${}^2\mathrm{F}_{7/2}(F=3)$}) of a single ${}^{171}$Yb$^+$ ion trapped in a radio-frequency (RF) Paul trap.
Detailed descriptions are given in Refs.~\cite{hun16} and \cite{san19}.
Its systematic frequency uncertainty was $u(\nu_\mathrm{Yb1E3})/\nu_\mathrm{Yb1E3} = 2.7 \times 10^{-18}$ during the measurements reported here.
Its clock laser operates at the first subharmonic of the clock transition and is pre-stabilized via a frequency comb to an ultrastable cryogenic silicon cavity \cite{mat17a}.
The resulting frequency instability of the reference clock is approximately $1 \times 10^{-15}/ \sqrt{\tau/\mathrm{s}}$ \cite{san19}.

\paragraph{Interferometric fibre link}
Interferometric fibre links \cite{pre12} enable phase-coherent optical frequency transfer across large distances using the telecom C Band ($\nu \approx 194~\mathrm{THz}$).
For this work, we set up a fibre link between PTB and MPQ that branches off the existing link between PTB and Strasbourg \cite{lis16}.
It follows the same route as in our previous chronometric levelling measurement \cite{gro24}.
Optical losses are compensated by a total of four fibre Brillouin amplifiers \cite{rau15} and one bidirectional erbium-doped fibre amplifier along the link.
The fibre link's contributions to the statistical and systematic uncertainties of the clock comparisons are negligible, where the accuracy is verified at a level below $1 \times 10^{-18}$ by round-trip measurements.
While frequency transfer across fibre links is subject to relativistic effects \cite{ger15a}, the resulting frequency shifts are at the level of few $10^{-20}$ and therefore can be neglected safely here.

\paragraph{Optical frequency combs}
At each station, an optical frequency comb connects the respective clock laser to the fibre link by measuring their optical frequency ratio.
At MPQ, the comb is housed in the same air-conditioned car trailer as the transportable clock.
At PTB, the comb is situated in a different building from the reference clock.
The contributions by frequency transfer at the combs to the statistical and systematic errors of the clock comparisons are negligible, where the accuracy has been verified for the comb at PTB at a level below $1 \times 10^{-18}$ by round-trip measurements via an independent comb \cite{ben19}.

\paragraph{Data recording and analysis}
We record and analyse data following the formalism of Lodewyck \textit{et al.} for optical clock networks \cite{lod20}.
The frequencies of all relevant beat notes are recorded by dead time--free counters, which are operated in phase-averaging mode ($\Lambda$) and with gate times of 1~s.
The counters at PTB and MPQ are synchronized to UTC(PTB) and a GPS-disciplined local timescale, respectively.
Stable RF references are derived from local hydrogen masers, but are not relevant for the clock comparison as we measure optical frequency ratios, which are dimensionless quantities, throughout.

The frequency ratio of the clocks is computed as the mean value of all valid data.
Here, the statistical uncertainty is determined by extrapolating the white frequency noise--limited instability of the comparison, as estimated from the data's overlapping Allan deviation, to the full length of the data set.

\subsection*{Modelling of correlations between frequency ratios}
In order to produce the best possible estimate of the uncertainty of the measured RRS $\Delta \nu / \nu$, we take into account the autocorrelation \cite{gum08} of each clock's systematic bias between the local \cite{vis26} and remote measurements.
In the following, these are described by autocorrelation coefficients $\rho$ for each clock, while cross-correlations between the clocks are negligible.

The systematic bias of the reference clock remains highly autocorrelated over time.
The largest uncorrelated contributions stems from the probe light shift cancellation servo, variations of the shifts from blackbody radiation and micromotion, estimated to be $0.5 \times 10^{-18}$ in total for each measurement.
Together with the systematic uncertainty of $2.7 \times 10^{-18}$, this results in $\rho_\mathrm{Yb1E3} = 0.97$.

The systematic bias of the transportable clock is partially autocorrelated over time:
Uncorrelated effects include the axial temperature of the atoms in the optical lattice, the depth of the optical lattice, the density shift and the clock laser servo error of the clock laser stabilization.
The blackbody radiation shift, second-order Zeeman shift and frequency shift due to collisions with background gas are each treated as fully correlated over time.
Overall, we find $\rho_\mathrm{Sr4} = 0.63$.

Table~\ref{tab:UncertaintyContributions} summarizes the contributions to uncertainties of the local and remote measurements as well as their difference.

\begin{table}[tbp]
	\caption{
        \label{tab:UncertaintyContributions}
        Uncertainty contributions of statistical noise, systematic biases of the clocks, and local levelling to the local \cite{vis26} and remote frequency ratio measurements as well as their difference $\Delta R$.
        All values are given in multiples of $10^{-18}$.
    }
	\begin{tabular}{llll}
        \hline
        Contribution    & $u(R_\mathrm{PTB}) / R$  & $u(R_\mathrm{MPQ}) / R$  & $u(\Delta R) / R$ \\
        \hline
        Stat.              & 2.1  & 3.2 & 3.8 \\
        Sys., Sr4 (sense)  & 2.6  & 3.3 & 2.6 \\
        Sys., Yb1E3 (ref.) & 2.7  & 2.7 & 0.7 \\
        Levelling          & 0.5  & --- & 0.5 \\
        \hline
        Total              & 4.3  & 5.3 & 4.7
	\end{tabular}
\end{table}

\subsection*{Enhancement of the GRS test and influence of the centrifugal potential}
In this work and similar GRS tests, the gravity potential difference $\Delta W$ and centrifugal potential difference $\Delta Z$ are independent input quantities, whereas the gravitational potential difference $\Delta U$ is a dependent quantity that is determined by these two through Eq.~(\ref{eq:gravity-potential}).
Solving Eq.~(\ref{eq:RelTest}) for $\alpha$ and inserting Eq.~(\ref{eq:gravity-potential}), the GRS violation parameter is given by
\begin{equation}
    \alpha =
    \frac{
        c^2 \frac{{\Delta\nu}}{\nu} - {\Delta W} - \alpha_\mathrm{SR} {\Delta Z}
    }{
        {\Delta W} - {\Delta Z}
    }
    \label{eq:alpha}
\end{equation}
in terms of those independent quantities.
Here, we investigate the influence of the enhancement factor $\eta = {\Delta U} / {\Delta W}$ and the centrifugal potential difference $\Delta Z$ on the GRS test's uncertainty $u(\alpha)$ for the geographic configuration discussed in the main text (${\Delta Z} / {\Delta W} < 0$), by applying linear error propagation to Eq.~(\ref{eq:alpha}) and considering that $|\alpha_\mathrm{SR}| \ll 1$.
The uncertainty contribution $u(\alpha_{\alpha_\mathrm{SR}})$ is small because $u(\alpha_{\alpha_\mathrm{SR}}) < u(\alpha_\mathrm{SR}) \ll u(\alpha)$.

The influence of the centrifugal potential difference's uncertainty $u(\Delta Z / c^2)$ on the GRS test's uncertainty $u(\alpha)$ is marginal, as it enters mainly through the denominator of Eq.~(\ref{eq:alpha}) because $|\alpha_\mathrm{SR}| \ll 1$.
The corresponding uncertainty contribution is given by
\begin{equation}
    u(\alpha_{{\Delta Z} / c^2})
    =       c^2 \left|\frac{\partial\alpha}{\partial{\Delta Z}}\right|
            u({\Delta Z} / c^2)
    \approx |\alpha| \frac{u({\Delta Z} / c^2)}{|\Delta U / c^2|}
    <       |\alpha| \frac{u({\Delta Z} / c^2)}{|\Delta Z / c^2|}
    \mbox{,}
\end{equation}
since $|\Delta U| > |\Delta Z|$ here.
Considering that $|\alpha| \lesssim u(\alpha)$, the uncertainty contribution is therefore negligible if $u({\Delta Z}/c^2) \ll |\Delta Z| / c^2$, which is typically the case.

The uncertainty contributions
\begin{equation}
    u(\alpha_{\Delta\nu/\nu})
    =       \eta^{-1} \frac{u({\Delta \nu} / \nu)}{|{\Delta W} / c^2|}
    \approx \eta^{-1} \frac{u({\Delta \nu} / \nu)}{|{\Delta \nu} / \nu|}
\end{equation}
from the clock comparisons and
\begin{equation}
    u(\alpha_{\Delta W/c^2})
    =       \eta^{-1} | 1 + \alpha | \frac{u({\Delta W/c^2})}{|{\Delta W} / c^2|}
    \approx \eta^{-1} \frac{u({\Delta W/c^2})}{|{\Delta W} / c^2|}
\end{equation}
from the determination of the gravity potential difference both reduce approximately by the enhancement factor $\eta$ compared to a configuration with $\Delta Z = 0$ but with otherwise identical parameters.
Since the uncertainty contributions from $\alpha_\mathrm{SR}$ and $\Delta Z$ are negligible, this results directly in an enhancement of the GRS test's sensitivity by the factor $\eta$ without increasing the uncertainty, as stated in the main text.

%
%
\bmhead{Acknowledgements}
%
We thank Chun Yu Ma, Mona Kempkes, Thomas Legero and Uwe Sterr for operating the ultrastable \qty{1.5}{\micro\meter} laser oscillator at PTB;
Kilian Stahl for substantial contribution to the preparation of Fig.~\ref{fig:schematic};
and
Serdar Seng\"ul for assistance with geometric levelling at MPQ.

\section*{Declarations}
%
\begin{itemize}
    \item \textbf{Funding}\\
    This work has been supported by
    the projects 22IEM01 TOCK,
                 23FUN02 CoCoRICO and
                 25IEM09 Arcsecond
        which have received funding from the European Partnership on Metrology, co-financed from the European Union’s Horizon Europe Research and Innovation Programme and by the Participating States;
    the Deutsche Forschungsgemeinschaft (DFG, German Research Foundation) under
        Germany’s Excellence Strategy -- EXC-2123 and EXC-2123/2 QuantumFrontiers -- Project-ID 390837967,
        SFB~1227 DQ-\textit{mat} -- Project-ID 274200144 -- within project B02
        and
        Research Unit FOR5456 -- Project-ID 490990195;
    and
    the Max Planck--RIKEN--PTB Center for Time, Constants and Fundamental Symmetries.
    \item \textbf{Competing interests}\\
    The authors declare no competing interests.
    \item \textbf{Ethics approval and consent to participate}\\
    Not applicable
    \item \textbf{Consent for publication}\\
    Not applicable
    \item \textbf{Data availability}\\
    The data supporting the findings of this study are available in the PTB Open Access Repository [DOI to be inserted].
    \item \textbf{Materials availability}\\
    Not applicable
    \item \textbf{Code availability }\\
    The code used to analyze the data in this study is available from the corresponding author upon reasonable request.
    \item \textbf{Author contribution}\\
    \IngoNosske, \ChetanVishwakarma and \TimLuecke operated the transportable optical clock, including its frequency comb, at MPQ.
    \MelinaFilzinger, \MartinSteinel and \NilsHuntemann operated the reference optical clock at PTB.
    \JochenKronjaeger, \AlexanderKuhl and \ShamboMukherjee operated the optical fibre link between PTB and MPQ.
    \ErikBenkler operated the frequency comb at PTB.
    \HeinerDenker determined the geopotential values, including time-variable parts (tides), of the height reference markers at both stations.
    \ChetanVishwakarma, \IngoNosske and \SoerenDoerscher, with support from all authors, wrote the manuscript.
    All authors contributed to the discussion and interpretation of the results.
\end{itemize}
%

\begin{thebibliography}{55}
\ifx \bisbn   \undefined \def \bisbn  #1{ISBN #1}\fi
\ifx \binits  \undefined \def \binits#1{#1}\fi
\ifx \bauthor  \undefined \def \bauthor#1{#1}\fi
\ifx \batitle  \undefined \def \batitle#1{#1}\fi
\ifx \bjtitle  \undefined \def \bjtitle#1{#1}\fi
\ifx \bvolume  \undefined \def \bvolume#1{\textbf{#1}}\fi
\ifx \byear  \undefined \def \byear#1{#1}\fi
\ifx \bissue  \undefined \def \bissue#1{#1}\fi
\ifx \bfpage  \undefined \def \bfpage#1{#1}\fi
\ifx \blpage  \undefined \def \blpage #1{#1}\fi
\ifx \burl  \undefined \def \burl#1{\textsf{#1}}\fi
\ifx \doiurl  \undefined \def \doiurl#1{\url{https://doi.org/#1}}\fi
\ifx \betal  \undefined \def \betal{\textit{et al.}}\fi
\ifx \binstitute  \undefined \def \binstitute#1{#1}\fi
\ifx \binstitutionaled  \undefined \def \binstitutionaled#1{#1}\fi
\ifx \bctitle  \undefined \def \bctitle#1{#1}\fi
\ifx \beditor  \undefined \def \beditor#1{#1}\fi
\ifx \bpublisher  \undefined \def \bpublisher#1{#1}\fi
\ifx \bbtitle  \undefined \def \bbtitle#1{#1}\fi
\ifx \bedition  \undefined \def \bedition#1{#1}\fi
\ifx \bseriesno  \undefined \def \bseriesno#1{#1}\fi
\ifx \blocation  \undefined \def \blocation#1{#1}\fi
\ifx \bsertitle  \undefined \def \bsertitle#1{#1}\fi
\ifx \bsnm \undefined \def \bsnm#1{#1}\fi
\ifx \bsuffix \undefined \def \bsuffix#1{#1}\fi
\ifx \bparticle \undefined \def \bparticle#1{#1}\fi
\ifx \barticle \undefined \def \barticle#1{#1}\fi
\bibcommenthead
\ifx \bconfdate \undefined \def \bconfdate #1{#1}\fi
\ifx \botherref \undefined \def \botherref #1{#1}\fi
\ifx \url \undefined \def \url#1{\textsf{#1}}\fi
\ifx \bchapter \undefined \def \bchapter#1{#1}\fi
\ifx \bbook \undefined \def \bbook#1{#1}\fi
\ifx \bcomment \undefined \def \bcomment#1{#1}\fi
\ifx \oauthor \undefined \def \oauthor#1{#1}\fi
\ifx \citeauthoryear \undefined \def \citeauthoryear#1{#1}\fi
\ifx \endbibitem  \undefined \def \endbibitem {}\fi
\ifx \bconflocation  \undefined \def \bconflocation#1{#1}\fi
\ifx \arxivurl  \undefined \def \arxivurl#1{\textsf{#1}}\fi
\csname PreBibitemsHook\endcsname

\bibitem[\protect\citeauthoryear{Einstein}{1916}]{ein16}
\begin{barticle}
\bauthor{\bsnm{Einstein}, \binits{A.}}:
\batitle{Die {G}rundlage der {A}llgemeinen {R}elativit\"atstheorie}.
\bjtitle{Annalen der Physik}
\bvolume{IV 49}(\bissue{7}),
\bfpage{769}--\blpage{822}
(\byear{1916})
\doiurl{10.1002/andp.19163540702}
\end{barticle}
\endbibitem

\bibitem[\protect\citeauthoryear{Will}{2014}]{wil14}
\begin{barticle}
\bauthor{\bsnm{Will}, \binits{C.M.}}:
\batitle{The confrontation between general relativity and experiment}.
\bjtitle{Living Rev. Relativity}
\bvolume{9},
\bfpage{3}
(\byear{2014})
\doiurl{10.12942/lrr-2014-4} .
\bcomment{[Online Article]: cited 2014-07-01, updated in 2014}
\end{barticle}
\endbibitem

\bibitem[\protect\citeauthoryear{Bertone and Hooper}{2018}]{ber18e}
\begin{barticle}
\bauthor{\bsnm{Bertone}, \binits{G.}},
\bauthor{\bsnm{Hooper}, \binits{D.}}:
\batitle{History of dark matter}.
\bjtitle{Rev. Mod. Phys.}
\bvolume{90},
\bfpage{045002}
(\byear{2018})
\doiurl{10.1103/RevModPhys.90.045002}
\end{barticle}
\endbibitem

\bibitem[\protect\citeauthoryear{Brax}{2017}]{bra17a}
\begin{barticle}
\bauthor{\bsnm{Brax}, \binits{P.}}:
\batitle{What makes the {U}niverse accelerate? {A} review on what dark energy could be and how to test it}.
\bjtitle{Rep. Prog. Phys.}
\bvolume{81}(\bissue{1}),
\bfpage{016902}
(\byear{2017})
\doiurl{10.1088/1361-6633/aa8e64}
\end{barticle}
\endbibitem

\bibitem[\protect\citeauthoryear{Bahcall et~al.}{1999}]{bah99}
\begin{barticle}
\bauthor{\bsnm{Bahcall}, \binits{N.A.}},
\bauthor{\bsnm{Ostriker}, \binits{J.P.}},
\bauthor{\bsnm{Perlmutter}, \binits{S.}},
\bauthor{\bsnm{Steinhardt}, \binits{P.J.}}:
\batitle{The cosmic triangle: Revealing the state of the universe}.
\bjtitle{Science}
\bvolume{284}(\bissue{5419}),
\bfpage{1481}--\blpage{1488}
(\byear{1999})
\doiurl{10.1126/science.284.5419.1481}
\end{barticle}
\endbibitem

\bibitem[\protect\citeauthoryear{Damour}{1999}]{dam99}
\begin{bchapter}
\bauthor{\bsnm{Damour}, \binits{T.}}:
\bctitle{Equivalence principle and clocks}.
In: \beditor{\bsnm{Tran Thanh~Van}, \binits{J.}},
\beditor{\bsnm{Dumarchez}, \binits{J.}},
\beditor{\bsnm{Reynaud}, \binits{S.}},
\beditor{\bsnm{Salomon}, \binits{C.}},
\beditor{\bsnm{Thorsett}, \binits{S.}} (eds.)
\bbtitle{34th Rencontres de {M}oriond: {G}ravitational Waves and Experimental Gravity},
pp. \bfpage{357}--\blpage{365}
(\byear{1999}).
\burl{https://cds.cern.ch/record/371944}
\end{bchapter}
\endbibitem

\bibitem[\protect\citeauthoryear{Delva et~al.}{2018}]{del18}
\begin{barticle}
\bauthor{\bsnm{Delva}, \binits{P.}},
\bauthor{\bsnm{Puchades}, \binits{N.}},
\bauthor{\bsnm{Sch\"onemann}, \binits{E.}},
\bauthor{\bsnm{Dilssner}, \binits{F.}},
\bauthor{\bsnm{Courde}, \binits{C.}},
\bauthor{\bsnm{Bertone}, \binits{S.}},
\bauthor{\bsnm{Gonzalez}, \binits{F.}},
\bauthor{\bsnm{Hees}, \binits{A.}},
\bauthor{\bsnm{Le~Poncin-Lafitte}, \binits{C.}},
\bauthor{\bsnm{Meynadier}, \binits{F.}},
\bauthor{\bsnm{Prieto-Cerdeira}, \binits{R.}},
\bauthor{\bsnm{Sohet}, \binits{B.}},
\bauthor{\bsnm{Ventura-Traveset}, \binits{J.}},
\bauthor{\bsnm{Wolf}, \binits{P.}}:
\batitle{Gravitational redshift test using eccentric galileo satellites}.
\bjtitle{Phys. Rev. Lett.}
\bvolume{121},
\bfpage{231101}
(\byear{2018})
\doiurl{10.1103/PhysRevLett.121.231101}
\end{barticle}
\endbibitem

\bibitem[\protect\citeauthoryear{Herrmann et~al.}{2018}]{her18b}
\begin{barticle}
\bauthor{\bsnm{Herrmann}, \binits{S.}},
\bauthor{\bsnm{Finke}, \binits{F.}},
\bauthor{\bsnm{L\"ulf}, \binits{M.}},
\bauthor{\bsnm{Kichakova}, \binits{O.}},
\bauthor{\bsnm{Puetzfeld}, \binits{D.}},
\bauthor{\bsnm{Knickmann}, \binits{D.}},
\bauthor{\bsnm{List}, \binits{M.}},
\bauthor{\bsnm{Rievers}, \binits{B.}},
\bauthor{\bsnm{Giorgi}, \binits{G.}},
\bauthor{\bsnm{G\"unther}, \binits{C.}},
\bauthor{\bsnm{Dittus}, \binits{H.}},
\bauthor{\bsnm{Prieto-Cerdeira}, \binits{R.}},
\bauthor{\bsnm{Dilssner}, \binits{F.}},
\bauthor{\bsnm{Gonzalez}, \binits{F.}},
\bauthor{\bsnm{Sch\"onemann}, \binits{E.}},
\bauthor{\bsnm{Ventura-Traveset}, \binits{J.}},
\bauthor{\bsnm{L\"ammerzahl}, \binits{C.}}:
\batitle{Test of the gravitational redshift with {G}alileo satellites in an eccentric orbit}.
\bjtitle{Phys. Rev. Lett.}
\bvolume{121},
\bfpage{231102}
(\byear{2018})
\doiurl{10.1103/PhysRevLett.121.231102}
\end{barticle}
\endbibitem

\bibitem[\protect\citeauthoryear{Takamoto et~al.}{2020}]{tak20}
\begin{barticle}
\bauthor{\bsnm{Takamoto}, \binits{M.}},
\bauthor{\bsnm{Ushijima}, \binits{I.}},
\bauthor{\bsnm{Ohmae}, \binits{N.}},
\bauthor{\bsnm{Yahagi}, \binits{T.}},
\bauthor{\bsnm{Kokado}, \binits{K.}},
\bauthor{\bsnm{Shinkai}, \binits{H.}},
\bauthor{\bsnm{Katori}, \binits{H.}}:
\batitle{Test of general relativity by a pair of transportable optical lattice clocks}.
\bjtitle{Nature Photonics}
\bvolume{14},
\bfpage{411}--\blpage{415}
(\byear{2020})
\doiurl{10.1038/s41566-020-0619-8}
\end{barticle}
\endbibitem

\bibitem[\protect\citeauthoryear{Pound and Rebka}{1960}]{pou60}
\begin{barticle}
\bauthor{\bsnm{Pound}, \binits{R.V.}},
\bauthor{\bsnm{Rebka}, \binits{G.A.}}:
\batitle{Apparent weight of photons}.
\bjtitle{Phys. Rev. Lett.}
\bvolume{4},
\bfpage{337}--\blpage{341}
(\byear{1960})
\doiurl{10.1103/PhysRevLett.4.337}
\end{barticle}
\endbibitem

\bibitem[\protect\citeauthoryear{Pound and Snider}{1965}]{pou65}
\begin{barticle}
\bauthor{\bsnm{Pound}, \binits{R.V.}},
\bauthor{\bsnm{Snider}, \binits{J.L.}}:
\batitle{Effect of gravity on gamma radiation}.
\bjtitle{Phys. Rev.}
\bvolume{140},
\bfpage{788}--\blpage{803}
(\byear{1965})
\doiurl{10.1103/PhysRev.140.B788}
\end{barticle}
\endbibitem

\bibitem[\protect\citeauthoryear{Takano et~al.}{2016}]{tak16}
\begin{barticle}
\bauthor{\bsnm{Takano}, \binits{T.}},
\bauthor{\bsnm{Takamoto}, \binits{M.}},
\bauthor{\bsnm{Ushijima}, \binits{I.}},
\bauthor{\bsnm{Ohmae}, \binits{N.}},
\bauthor{\bsnm{Akatsuka}, \binits{T.}},
\bauthor{\bsnm{Yamaguchi}, \binits{A.}},
\bauthor{\bsnm{Kuroishi}, \binits{Y.}},
\bauthor{\bsnm{Munekane}, \binits{H.}},
\bauthor{\bsnm{Miyahara}, \binits{B.}},
\bauthor{\bsnm{Katori}, \binits{H.}}:
\batitle{Geopotential measurements with synchronously linked optical lattice clocks}.
\bjtitle{Nature Photonics}
\bvolume{10},
\bfpage{662}--\blpage{666}
(\byear{2016})
\doiurl{10.1038/nphoton.2016.159}
\end{barticle}
\endbibitem

\bibitem[\protect\citeauthoryear{Grotti et~al.}{2018}]{gro18a}
\begin{barticle}
\bauthor{\bsnm{Grotti}, \binits{J.}},
\bauthor{\bsnm{Koller}, \binits{S.}},
\bauthor{\bsnm{Vogt}, \binits{S.}},
\bauthor{\bsnm{H\"afner}, \binits{S.}},
\bauthor{\bsnm{Sterr}, \binits{U.}},
\bauthor{\bsnm{Lisdat}, \binits{C.}},
\bauthor{\bsnm{Denker}, \binits{H.}},
\bauthor{\bsnm{Voigt}, \binits{C.}},
\bauthor{\bsnm{Timmen}, \binits{L.}},
\bauthor{\bsnm{Rolland}, \binits{A.}},
\bauthor{\bsnm{Baynes}, \binits{F.N.}},
\bauthor{\bsnm{Margolis}, \binits{H.S.}},
\bauthor{\bsnm{Zampaolo}, \binits{M.}},
\bauthor{\bsnm{Thoumany}, \binits{P.}},
\bauthor{\bsnm{Pizzocaro}, \binits{M.}},
\bauthor{\bsnm{Rauf}, \binits{B.}},
\bauthor{\bsnm{Bregolin}, \binits{F.}},
\bauthor{\bsnm{Tampellini}, \binits{A.}},
\bauthor{\bsnm{Barbieri}, \binits{P.}},
\bauthor{\bsnm{Zucco}, \binits{M.}},
\bauthor{\bsnm{Costanzo}, \binits{G.A.}},
\bauthor{\bsnm{Clivati}, \binits{C.}},
\bauthor{\bsnm{Levi}, \binits{F.}},
\bauthor{\bsnm{Calonico}, \binits{D.}}:
\batitle{Geodesy and metrology with a transportable optical clock}.
\bjtitle{Nature Physics}
\bvolume{14},
\bfpage{437}--\blpage{441}
(\byear{2018})
\doiurl{10.1038/s41567-017-0042-3}
\end{barticle}
\endbibitem

\bibitem[\protect\citeauthoryear{Smarr et~al.}{1983}]{sma83}
\begin{barticle}
\bauthor{\bsnm{Smarr}, \binits{L.L.}},
\bauthor{\bsnm{Vessot}, \binits{R.F.C.}},
\bauthor{\bsnm{Lundquist}, \binits{C.A.}},
\bauthor{\bsnm{Decher}, \binits{R.}},
\bauthor{\bsnm{Piran}, \binits{T.}}:
\batitle{Gravitational waves and red shifts: {A} space experiment for testing relativistic gravity using multiple time-correlated radio signals}.
\bjtitle{Gen. Rel.Grav.}
\bvolume{15}(\bissue{2}),
\bfpage{129}--\blpage{163}
(\byear{1983})
\doiurl{10.1007/BF00762473}
\end{barticle}
\endbibitem

\bibitem[\protect\citeauthoryear{De~Marchi et~al.}{2023}]{mar23}
\begin{barticle}
\bauthor{\bsnm{De~Marchi}, \binits{F.}},
\bauthor{\bsnm{Cascioli}, \binits{G.}},
\bauthor{\bsnm{Ely}, \binits{T.}},
\bauthor{\bsnm{Iess}, \binits{L.}},
\bauthor{\bsnm{Burt}, \binits{E.A.}},
\bauthor{\bsnm{Hensley}, \binits{S.}},
\bauthor{\bsnm{Mazarico}, \binits{E.}}:
\batitle{Testing the gravitational redshift with an inner {Solar} {System} probe: {The} {VERITAS} case}.
\bjtitle{Phys. Rev. D}
\bvolume{107},
\bfpage{064032}
(\byear{2023})
\doiurl{10.1103/PhysRevD.107.064032}
\end{barticle}
\endbibitem

\bibitem[\protect\citeauthoryear{Delva et~al.}{2017}]{del17}
\begin{barticle}
\bauthor{\bsnm{Delva}, \binits{P.}},
\bauthor{\bsnm{Lodewyck}, \binits{J.}},
\bauthor{\bsnm{Bilicki}, \binits{S.}},
\bauthor{\bsnm{Bookjans}, \binits{E.}},
\bauthor{\bsnm{Vallet}, \binits{G.}},
\bauthor{\bsnm{Le~Targat}, \binits{R.}},
\bauthor{\bsnm{Pottie}, \binits{P.-E.}},
\bauthor{\bsnm{Guerlin}, \binits{C.}},
\bauthor{\bsnm{Meynadier}, \binits{F.}},
\bauthor{\bsnm{Le~Poncin-Lafitte}, \binits{C.}},
\bauthor{\bsnm{Lopez}, \binits{O.}},
\bauthor{\bsnm{Amy-Klein}, \binits{A.}},
\bauthor{\bsnm{Lee}, \binits{W.-K.}},
\bauthor{\bsnm{Quintin}, \binits{N.}},
\bauthor{\bsnm{Lisdat}, \binits{C.}},
\bauthor{\bsnm{Al-Masoudi}, \binits{A.}},
\bauthor{\bsnm{D\"orscher}, \binits{S.}},
\bauthor{\bsnm{Grebing}, \binits{C.}},
\bauthor{\bsnm{Grosche}, \binits{G.}},
\bauthor{\bsnm{Kuhl}, \binits{A.}},
\bauthor{\bsnm{Raupach}, \binits{S.}},
\bauthor{\bsnm{Sterr}, \binits{U.}},
\bauthor{\bsnm{Hill}, \binits{I.R.}},
\bauthor{\bsnm{Hobson}, \binits{R.}},
\bauthor{\bsnm{Bowden}, \binits{W.}},
\bauthor{\bsnm{Kronj\"ager}, \binits{J.}},
\bauthor{\bsnm{Marra}, \binits{G.}},
\bauthor{\bsnm{Rolland}, \binits{A.}},
\bauthor{\bsnm{Baynes}, \binits{F.N.}},
\bauthor{\bsnm{Margolis}, \binits{H.S.}},
\bauthor{\bsnm{Gill}, \binits{P.}}:
\batitle{Test of special relativity using a fiber network of optical clocks}.
\bjtitle{Phys. Rev. Lett.}
\bvolume{118},
\bfpage{221102}
(\byear{2017})
\doiurl{10.1103/PhysRevLett.118.221102}
\end{barticle}
\endbibitem

\bibitem[\protect\citeauthoryear{Denker et~al.}{2018}]{den17}
\begin{barticle}
\bauthor{\bsnm{Denker}, \binits{H.}},
\bauthor{\bsnm{Timmen}, \binits{L.}},
\bauthor{\bsnm{Voigt}, \binits{C.}},
\bauthor{\bsnm{Weyers}, \binits{S.}},
\bauthor{\bsnm{Peik}, \binits{E.}},
\bauthor{\bsnm{Margolis}, \binits{H.S.}},
\bauthor{\bsnm{Delva}, \binits{P.}},
\bauthor{\bsnm{Wolf}, \binits{P.}},
\bauthor{\bsnm{Petit}, \binits{G.}}:
\batitle{Geodetic methods to determine the relativistic redshift at the level of $10^{-18}$ in the context of international timescales: a review and practical results}.
\bjtitle{J. Geod.}
\bvolume{92},
\bfpage{487}--\blpage{516}
(\byear{2018})
\doiurl{10.1007/s00190-017-1075-1}
\end{barticle}
\endbibitem

\bibitem[\protect\citeauthoryear{Riedel et~al.}{2020}]{rie20}
\begin{barticle}
\bauthor{\bsnm{Riedel}, \binits{F.}},
\bauthor{\bsnm{Al-Masoudi}, \binits{A.}},
\bauthor{\bsnm{Benkler}, \binits{E.}},
\bauthor{\bsnm{D\"orscher}, \binits{S.}},
\bauthor{\bsnm{Gerginov}, \binits{V.}},
\bauthor{\bsnm{Grebing}, \binits{C.}},
\bauthor{\bsnm{H\"afner}, \binits{S.}},
\bauthor{\bsnm{Huntemann}, \binits{N.}},
\bauthor{\bsnm{Lipphardt}, \binits{B.}},
\bauthor{\bsnm{Lisdat}, \binits{C.}},
\bauthor{\bsnm{Peik}, \binits{E.}},
\bauthor{\bsnm{Piester}, \binits{D.}},
\bauthor{\bsnm{Sanner}, \binits{C.}},
\bauthor{\bsnm{Tamm}, \binits{C.}},
\bauthor{\bsnm{Weyers}, \binits{S.}},
\bauthor{\bsnm{Denker}, \binits{H.}},
\bauthor{\bsnm{Timmen}, \binits{L.}},
\bauthor{\bsnm{Voigt}, \binits{C.}},
\bauthor{\bsnm{Calonico}, \binits{D.}},
\bauthor{\bsnm{Cerretto}, \binits{G.}},
\bauthor{\bsnm{Costanzo}, \binits{G.A.}},
\bauthor{\bsnm{Levi}, \binits{F.}},
\bauthor{\bsnm{Sesia}, \binits{I.}},
\bauthor{\bsnm{Achkar}, \binits{J.}},
\bauthor{\bsnm{Gu\'ena}, \binits{J.}},
\bauthor{\bsnm{Abgrall}, \binits{M.}},
\bauthor{\bsnm{Rovera}, \binits{G.D.}},
\bauthor{\bsnm{Chupin}, \binits{B.}},
\bauthor{\bsnm{Shi}, \binits{C.}},
\bauthor{\bsnm{Bilicki}, \binits{S.}},
\bauthor{\bsnm{Bookjans}, \binits{E.}},
\bauthor{\bsnm{Lodewyck}, \binits{J.}},
\bauthor{\bsnm{Le~Targat}, \binits{R.}},
\bauthor{\bsnm{Delva}, \binits{P.}},
\bauthor{\bsnm{Bize}, \binits{S.}},
\bauthor{\bsnm{Baynes}, \binits{F.N.}},
\bauthor{\bsnm{Baynham}, \binits{C.}},
\bauthor{\bsnm{Bowden}, \binits{W.}},
\bauthor{\bsnm{Gill}, \binits{P.}},
\bauthor{\bsnm{Godun}, \binits{R.M.}},
\bauthor{\bsnm{Hill}, \binits{I.R.}},
\bauthor{\bsnm{Hobson}, \binits{R.}},
\bauthor{\bsnm{Jones}, \binits{J.M.}},
\bauthor{\bsnm{King}, \binits{S.A.}},
\bauthor{\bsnm{Nisbet-Jones}, \binits{P.}},
\bauthor{\bsnm{Rolland}, \binits{A.}},
\bauthor{\bsnm{Shemar}, \binits{S.L.}},
\bauthor{\bsnm{Whibberley}, \binits{P.B.}},
\bauthor{\bsnm{Margolis}, \binits{H.S.}}:
\batitle{Direct comparisons of {E}uropean primary and secondary frequency standards via satellite techniques}.
\bjtitle{Metrologia}
\bvolume{57}(\bissue{4}),
\bfpage{045005}
(\byear{2020})
\doiurl{10.1088/1681-7575/ab6745}
\end{barticle}
\endbibitem

\bibitem[\protect\citeauthoryear{Nosske et~al.}{2025}]{nos25}
\begin{barticle}
\bauthor{\bsnm{Nosske}, \binits{I.}},
\bauthor{\bsnm{Vishwakarma}, \binits{C.}},
\bauthor{\bsnm{L\"ucke}, \binits{T.}},
\bauthor{\bsnm{Rahm}, \binits{J.}},
\bauthor{\bsnm{Poudel}, \binits{N.}},
\bauthor{\bsnm{Weyers}, \binits{S.}},
\bauthor{\bsnm{Benkler}, \binits{E.}},
\bauthor{\bsnm{D\"orscher}, \binits{S.}},
\bauthor{\bsnm{Lisdat}, \binits{C.}}:
\batitle{Transportable strontium lattice clock with $4 \times 10^{-19}$ blackbody radiation shift uncertainty}.
\bjtitle{Quantum Sci. Technol.}
\bvolume{10}(\bissue{4}),
\bfpage{045076}
(\byear{2025})
\doiurl{10.1088/2058-9565/ae1161}
\end{barticle}
\endbibitem

\bibitem[\protect\citeauthoryear{Huntemann et~al.}{2016}]{hun16}
\begin{barticle}
\bauthor{\bsnm{Huntemann}, \binits{N.}},
\bauthor{\bsnm{Sanner}, \binits{C.}},
\bauthor{\bsnm{Lipphardt}, \binits{B.}},
\bauthor{\bsnm{Tamm}, \binits{C.}},
\bauthor{\bsnm{Peik}, \binits{E.}}:
\batitle{Single-ion atomic clock with $3\times{10}^{-18}$ systematic uncertainty}.
\bjtitle{Phys. Rev. Lett.}
\bvolume{116},
\bfpage{063001}
(\byear{2016})
\doiurl{10.1103/PhysRevLett.116.063001}
\end{barticle}
\endbibitem

\bibitem[\protect\citeauthoryear{Sanner et~al.}{2019}]{san19}
\begin{barticle}
\bauthor{\bsnm{Sanner}, \binits{C.}},
\bauthor{\bsnm{Huntemann}, \binits{N.}},
\bauthor{\bsnm{Lange}, \binits{R.}},
\bauthor{\bsnm{Tamm}, \binits{C.}},
\bauthor{\bsnm{Peik}, \binits{E.}},
\bauthor{\bsnm{Safronova}, \binits{M.S.}},
\bauthor{\bsnm{Porsev}, \binits{S.G.}}:
\batitle{Optical clock comparison test of {L}orentz symmetry}.
\bjtitle{Nature}
\bvolume{567},
\bfpage{204}--\blpage{208}
(\byear{2019})
\doiurl{10.1038/s41586-019-0972-2}
\end{barticle}
\endbibitem

\bibitem[\protect\citeauthoryear{Vishwakarma et~al.}{2026}]{vis26}
\begin{botherref}
\oauthor{\bsnm{Vishwakarma}, \binits{C.}},
\oauthor{\bsnm{Nosske}, \binits{I.}},
\oauthor{\bsnm{L\"ucke}, \binits{T.}},
\oauthor{\bsnm{Steinel}, \binits{M.}},
\oauthor{\bsnm{Filzinger}, \binits{M.}},
\oauthor{\bsnm{Benkler}, \binits{E.}},
\oauthor{\bsnm{D\"orscher}, \binits{S.}},
\oauthor{\bsnm{Huntemann}, \binits{N.}},
\oauthor{\bsnm{Lisdat}, \binits{C.}}:
Interspecies clock comparison below $5 \times 10^{-18}$ uncertainty with a transportable clock.
arXiv:2608.01916 [physics.atom-ph]
(2026).
\doiurl{10.48550/arXiv.2608.01916}
\end{botherref}
\endbibitem

\bibitem[\protect\citeauthoryear{Riley}{2008}]{ril08}
\begin{botherref}
\oauthor{\bsnm{Riley}, \binits{W.J.}}:
Handbook of frequency stability analysis.
{NIST} special publication 1065,
NIST
(2008).
\url{http://tf.nist.gov/timefreq/general/pdf/2220.pdf}
\end{botherref}
\endbibitem

\bibitem[\protect\citeauthoryear{Greenhall and Riley}{2003}]{gre03b}
\begin{bchapter}
\bauthor{\bsnm{Greenhall}, \binits{C.}},
\bauthor{\bsnm{Riley}, \binits{W.}}:
\bctitle{Uncertainty of stability variances based on finite differences}.
In: \bbtitle{Proc. 2003 PTTI Meeting},
pp. \bfpage{267}--\blpage{280}
(\byear{2003}).
\burl{https://ntrs.nasa.gov/citations/20050061319}
\end{bchapter}
\endbibitem

\bibitem[\protect\citeauthoryear{Mohr and Taylor}{2000}]{moh00}
\begin{barticle}
\bauthor{\bsnm{Mohr}, \binits{P.J.}},
\bauthor{\bsnm{Taylor}, \binits{B.N.}}:
\batitle{{CODATA} recommended values of the fundamental physical constants: 1998}.
\bjtitle{Rev. Mod. Phys.}
\bvolume{72},
\bfpage{351}--\blpage{495}
(\byear{2000})
\doiurl{10.1103/RevModPhys.72.351}
\end{barticle}
\endbibitem

\bibitem[\protect\citeauthoryear{Grotti et~al.}{2024}]{gro24}
\begin{barticle}
\bauthor{\bsnm{Grotti}, \binits{J.}},
\bauthor{\bsnm{Nosske}, \binits{I.}},
\bauthor{\bsnm{Koller}, \binits{S.B.}},
\bauthor{\bsnm{Herbers}, \binits{S.}},
\bauthor{\bsnm{Denker}, \binits{H.}},
\bauthor{\bsnm{Timmen}, \binits{L.}},
\bauthor{\bsnm{Vishnyakova}, \binits{G.}},
\bauthor{\bsnm{Grosche}, \binits{G.}},
\bauthor{\bsnm{Waterholter}, \binits{T.}},
\bauthor{\bsnm{Kuhl}, \binits{A.}},
\bauthor{\bsnm{Koke}, \binits{S.}},
\bauthor{\bsnm{Benkler}, \binits{E.}},
\bauthor{\bsnm{Giunta}, \binits{M.}},
\bauthor{\bsnm{Maisenbacher}, \binits{L.}},
\bauthor{\bsnm{Matveev}, \binits{A.}},
\bauthor{\bsnm{D\"orscher}, \binits{S.}},
\bauthor{\bsnm{Schwarz}, \binits{R.}},
\bauthor{\bsnm{Al-Masoudi}, \binits{A.}},
\bauthor{\bsnm{H\"ansch}, \binits{T.W.}},
\bauthor{\bsnm{Udem}, \binits{T.}},
\bauthor{\bsnm{Holzwarth}, \binits{R.}},
\bauthor{\bsnm{Lisdat}, \binits{C.}}:
\batitle{Long-distance chronometric leveling with a portable optical clock}.
\bjtitle{Phys. Rev. Appl.}
\bvolume{21},
\bfpage{061001}
(\byear{2024})
\doiurl{10.1103/PhysRevApplied.21.L061001}
\end{barticle}
\endbibitem

\bibitem[\protect\citeauthoryear{Lisdat et~al.}{2016}]{lis16}
\begin{barticle}
\bauthor{\bsnm{Lisdat}, \binits{C.}},
\bauthor{\bsnm{Grosche}, \binits{G.}},
\bauthor{\bsnm{Quintin}, \binits{N.}},
\bauthor{\bsnm{Shi}, \binits{C.}},
\bauthor{\bsnm{Raupach}, \binits{S.M.F.}},
\bauthor{\bsnm{Grebing}, \binits{C.}},
\bauthor{\bsnm{Nicolodi}, \binits{D.}},
\bauthor{\bsnm{Stefani}, \binits{F.}},
\bauthor{\bsnm{Al-Masoudi}, \binits{A.}},
\bauthor{\bsnm{D\"orscher}, \binits{S.}},
\bauthor{\bsnm{H\"afner}, \binits{S.}},
\bauthor{\bsnm{Robyr}, \binits{J.-L.}},
\bauthor{\bsnm{Chiodo}, \binits{N.}},
\bauthor{\bsnm{Bilicki}, \binits{S.}},
\bauthor{\bsnm{Bookjans}, \binits{E.}},
\bauthor{\bsnm{Koczwara}, \binits{A.}},
\bauthor{\bsnm{Koke}, \binits{S.}},
\bauthor{\bsnm{Kuhl}, \binits{A.}},
\bauthor{\bsnm{Wiotte}, \binits{F.}},
\bauthor{\bsnm{Meynadier}, \binits{F.}},
\bauthor{\bsnm{Camisard}, \binits{E.}},
\bauthor{\bsnm{Abgrall}, \binits{M.}},
\bauthor{\bsnm{Lours}, \binits{M.}},
\bauthor{\bsnm{Legero}, \binits{T.}},
\bauthor{\bsnm{Schnatz}, \binits{H.}},
\bauthor{\bsnm{Sterr}, \binits{U.}},
\bauthor{\bsnm{Denker}, \binits{H.}},
\bauthor{\bsnm{Chardonnet}, \binits{C.}},
\bauthor{\bsnm{Le~Coq}, \binits{Y.}},
\bauthor{\bsnm{Santarelli}, \binits{G.}},
\bauthor{\bsnm{Amy-Klein}, \binits{A.}},
\bauthor{\bsnm{Le~Targat}, \binits{R.}},
\bauthor{\bsnm{Lodewyck}, \binits{J.}},
\bauthor{\bsnm{Lopez}, \binits{O.}},
\bauthor{\bsnm{Pottie}, \binits{P.-E.}}:
\batitle{A clock network for geodesy and fundamental science}.
\bjtitle{Nature Commun.}
\bvolume{7},
\bfpage{12443}
(\byear{2016})
\doiurl{10.1038/ncomms12443}
\end{barticle}
\endbibitem

\bibitem[\protect\citeauthoryear{Clivati et~al.}{2022}]{cli22b}
\begin{barticle}
\bauthor{\bsnm{Clivati}, \binits{C.}},
\bauthor{\bsnm{Pizzocaro}, \binits{M.}},
\bauthor{\bsnm{Bertacco}, \binits{E.K.}},
\bauthor{\bsnm{Condio}, \binits{S.}},
\bauthor{\bsnm{Costanzo}, \binits{G.A.}},
\bauthor{\bsnm{Donadello}, \binits{S.}},
\bauthor{\bsnm{Goti}, \binits{I.}},
\bauthor{\bsnm{Gozzelino}, \binits{M.}},
\bauthor{\bsnm{Levi}, \binits{F.}},
\bauthor{\bsnm{Mura}, \binits{A.}},
\bauthor{\bsnm{Risaro}, \binits{M.}},
\bauthor{\bsnm{Calonico}, \binits{D.}},
\bauthor{\bsnm{T\o{}nnes}, \binits{M.}},
\bauthor{\bsnm{Pointard}, \binits{B.}},
\bauthor{\bsnm{Mazouth-Laurol}, \binits{M.}},
\bauthor{\bsnm{Le~Targat}, \binits{R.}},
\bauthor{\bsnm{Abgrall}, \binits{M.}},
\bauthor{\bsnm{Lours}, \binits{M.}},
\bauthor{\bsnm{Le~Goff}, \binits{H.}},
\bauthor{\bsnm{Lorini}, \binits{L.}},
\bauthor{\bsnm{Pottie}, \binits{P.-E.}},
\bauthor{\bsnm{Cantin}, \binits{E.}},
\bauthor{\bsnm{Lopez}, \binits{O.}},
\bauthor{\bsnm{Chardonnet}, \binits{C.}},
\bauthor{\bsnm{Amy-Klein}, \binits{A.}}:
\batitle{Coherent optical-fiber link across {Italy} and {France}}.
\bjtitle{Phys. Rev. Appl.}
\bvolume{18},
\bfpage{054009}
(\byear{2022})
\doiurl{10.1103/PhysRevApplied.18.054009}
\end{barticle}
\endbibitem

\bibitem[\protect\citeauthoryear{Cantin et~al.}{2024}]{can24}
\begin{bchapter}
\bauthor{\bsnm{Cantin}, \binits{E.}},
\bauthor{\bsnm{Lopez}, \binits{O.}},
\bauthor{\bsnm{Chardonnet}, \binits{C.}},
\bauthor{\bsnm{Amy-Klein}, \binits{A.}},
\bauthor{\bsnm{Rabault}, \binits{M.}},
\bauthor{\bsnm{Coget}, \binits{G.}},
\bauthor{\bsnm{Rosenbusch}, \binits{P.}},
\bauthor{\bsnm{M{\'e}noret}, \binits{V.}},
\bauthor{\bsnm{Quintin}, \binits{N.}},
\bauthor{\bsnm{T{\o}nnes}, \binits{M.}}, \betal:
\bctitle{{REFIMEVE} frequency and time network and applications}.
In: \bbtitle{Journal of Physics: Conference Series},
vol. \bseriesno{2889},
p. \bfpage{012031}
(\byear{2024}).
\bcomment{IOP Publishing}
\end{bchapter}
\endbibitem

\bibitem[\protect\citeauthoryear{Ishibashi et~al.}{2025}]{ish25a}
\begin{barticle}
\bauthor{\bsnm{Ishibashi}, \binits{K.}},
\bauthor{\bsnm{Yorozu}, \binits{S.}},
\bauthor{\bsnm{Arima}, \binits{T.}},
\bauthor{\bsnm{Kawamura}, \binits{M.}},
\bauthor{\bsnm{Tokura}, \binits{Y.}},
\bauthor{\bsnm{Karube}, \binits{K.}},
\bauthor{\bsnm{Yu}, \binits{X.}},
\bauthor{\bsnm{Taguchi}, \binits{Y.}},
\bauthor{\bsnm{Hanaguri}, \binits{T.}},
\bauthor{\bsnm{Machida}, \binits{T.}},
\bauthor{\bsnm{Itahashi}, \binits{Y.M.}},
\bauthor{\bsnm{Iwasa}, \binits{Y.}},
\bauthor{\bsnm{Nishikawa}, \binits{H.}},
\bauthor{\bsnm{Araoka}, \binits{F.}},
\bauthor{\bsnm{Hioki}, \binits{T.}},
\bauthor{\bsnm{Saitoh}, \binits{E.}},
\bauthor{\bsnm{Deacon}, \binits{R.S.}},
\bauthor{\bsnm{Yamamoto}, \binits{M.}},
\bauthor{\bsnm{Fang}, \binits{N.}},
\bauthor{\bsnm{Kato}, \binits{Y.K.}},
\bauthor{\bsnm{Hida}, \binits{A.}},
\bauthor{\bsnm{Takamoto}, \binits{M.}},
\bauthor{\bsnm{Katori}, \binits{H.}},
\bauthor{\bsnm{L{\'e}s{\'e}leuc}, \binits{S.}},
\bauthor{\bsnm{Aoki}, \binits{T.}},
\bauthor{\bsnm{Yonezawa}, \binits{H.}},
\bauthor{\bsnm{Furusawa}, \binits{A.}},
\bauthor{\bsnm{Tabuchi}, \binits{Y.}},
\bauthor{\bsnm{Tamate}, \binits{S.}},
\bauthor{\bsnm{Abe}, \binits{E.}},
\bauthor{\bsnm{Nakamura}, \binits{Y.}},
\bauthor{\bsnm{Nakajima}, \binits{T.}},
\bauthor{\bsnm{Tarucha}, \binits{S.}},
\bauthor{\bsnm{Seki}, \binits{K.}},
\bauthor{\bsnm{Shirakawa}, \binits{T.}},
\bauthor{\bsnm{Yunoki}, \binits{S.}},
\bauthor{\bsnm{Nagaosa}, \binits{N.}}:
\batitle{Research on quantum materials and quantum technology at {RIKEN}}.
\bjtitle{ACS Nano}
\bvolume{19}(\bissue{13}),
\bfpage{12427}--\blpage{12457}
(\byear{2025})
\doiurl{10.1021/acsnano.4c15409}
\end{barticle}
\endbibitem

\bibitem[\protect\citeauthoryear{L{\"a}mmerzahl and Dittus}{2002}]{lam02}
\begin{barticle}
\bauthor{\bsnm{L{\"a}mmerzahl}, \binits{C.}},
\bauthor{\bsnm{Dittus}, \binits{H.}}:
\batitle{Fundamental physics in space: A guide to present projects}.
\bjtitle{Ann. Phys. (Leipzig)}
\bvolume{514}(\bissue{2}),
\bfpage{95}--\blpage{150}
(\byear{2002})
\doiurl{10.1002/andp.20025140201}
\end{barticle}
\endbibitem

\bibitem[\protect\citeauthoryear{Cacciapuoti and Salomon}{2009}]{cac09}
\begin{barticle}
\bauthor{\bsnm{Cacciapuoti}, \binits{L.}},
\bauthor{\bsnm{Salomon}, \binits{C.}}:
\batitle{Space clocks and fundamental tests: The {ACES} experiment}.
\bjtitle{Eur. Phys. J. Special Topics}
\bvolume{172},
\bfpage{57}--\blpage{68}
(\byear{2009})
\doiurl{10.1140/epjst/e2009-01041-7}
\end{barticle}
\endbibitem

\bibitem[\protect\citeauthoryear{Savalle et~al.}{2019}]{sav19a}
\begin{barticle}
\bauthor{\bsnm{Savalle}, \binits{E.}},
\bauthor{\bsnm{Guerlin}, \binits{C.}},
\bauthor{\bsnm{Delva}, \binits{P.}},
\bauthor{\bsnm{Meynadier}, \binits{F.}},
\bauthor{\bsnm{Poncin-Lafitte}, \binits{C.}},
\bauthor{\bsnm{Wolf}, \binits{P.}}:
\batitle{Gravitational redshift test with the future {ACES} mission}.
\bjtitle{Classical and Quantum Gravity}
\bvolume{36}(\bissue{24}),
\bfpage{245004}
(\byear{2019})
\doiurl{10.1088/1361-6382/ab4f25}
\end{barticle}
\endbibitem

\bibitem[\protect\citeauthoryear{McGrew et~al.}{2018}]{mcg18}
\begin{barticle}
\bauthor{\bsnm{McGrew}, \binits{W.F.}},
\bauthor{\bsnm{Zhang}, \binits{X.}},
\bauthor{\bsnm{Fasano}, \binits{R.J.}},
\bauthor{\bsnm{Sch\"affer}, \binits{S.A.}},
\bauthor{\bsnm{Beloy}, \binits{K.}},
\bauthor{\bsnm{Nicolodi}, \binits{D.}},
\bauthor{\bsnm{Brown}, \binits{R.C.}},
\bauthor{\bsnm{Hinkley}, \binits{N.}},
\bauthor{\bsnm{Milan}, \binits{G.}},
\bauthor{\bsnm{Schioppo}, \binits{M.}},
\bauthor{\bsnm{Yoon}, \binits{T.H.}},
\bauthor{\bsnm{Ludlow}, \binits{A.D.}}:
\batitle{Atomic clock performance beyond the geodetic limit}.
\bjtitle{Nature}
\bvolume{564},
\bfpage{87}--\blpage{90}
(\byear{2018})
\doiurl{10.1038/s41586-018-0738-2}
\end{barticle}
\endbibitem

\bibitem[\protect\citeauthoryear{Arnold et~al.}{2026}]{arn26a}
\begin{barticle}
\bauthor{\bsnm{Arnold}, \binits{K.J.}},
\bauthor{\bsnm{Lee}, \binits{M.D.K.}},
\bauthor{\bsnm{Zhao}, \binits{Q.}},
\bauthor{\bsnm{Qin}, \binits{Q.}},
\bauthor{\bsnm{Zhang}, \binits{Z.}},
\bauthor{\bsnm{Jayjong}, \binits{N.}},
\bauthor{\bsnm{Barrett}, \binits{M.D.}}:
\batitle{Lu\textsuperscript{+} optical frequency references with accuracy verified at the 19th digit}.
\bjtitle{Nature}
\bvolume{658},
\bfpage{60}
(\byear{2026})
\doiurl{10.1038/s41586-026-11072-8}
\end{barticle}
\endbibitem

\bibitem[\protect\citeauthoryear{Shen et~al.}{2022}]{she22}
\begin{barticle}
\bauthor{\bsnm{Shen}, \binits{Q.}},
\bauthor{\bsnm{Guan}, \binits{J.-Y.}},
\bauthor{\bsnm{Ren}, \binits{J.-G.}},
\bauthor{\bsnm{Zeng}, \binits{T.}},
\bauthor{\bsnm{Hou}, \binits{L.}},
\bauthor{\bsnm{Li}, \binits{M.}},
\bauthor{\bsnm{Cao}, \binits{Y.}},
\bauthor{\bsnm{Han}, \binits{J.-J.}},
\bauthor{\bsnm{Lian}, \binits{M.-Z.}},
\bauthor{\bsnm{Chen}, \binits{Y.-W.}},
\bauthor{\bsnm{Peng}, \binits{X.-X.}},
\bauthor{\bsnm{Wang}, \binits{S.-M.}},
\bauthor{\bsnm{Zhu}, \binits{D.-Y.}},
\bauthor{\bsnm{Shi}, \binits{X.-P.}},
\bauthor{\bsnm{Wang}, \binits{Z.-G.}},
\bauthor{\bsnm{Li}, \binits{Y.}},
\bauthor{\bsnm{Liu}, \binits{W.-Y.}},
\bauthor{\bsnm{Pan}, \binits{G.-S.}},
\bauthor{\bsnm{Wang}, \binits{Y.}},
\bauthor{\bsnm{Li}, \binits{Z.-H.}},
\bauthor{\bsnm{Wu}, \binits{J.-C.}},
\bauthor{\bsnm{Zhang}, \binits{Y.-Y.}},
\bauthor{\bsnm{Chen}, \binits{F.-X.}},
\bauthor{\bsnm{Lu}, \binits{C.-Y.}},
\bauthor{\bsnm{Liao}, \binits{S.-K.}},
\bauthor{\bsnm{Yin}, \binits{J.}},
\bauthor{\bsnm{Jia}, \binits{J.-J.}},
\bauthor{\bsnm{Peng}, \binits{C.-Z.}},
\bauthor{\bsnm{Jiang}, \binits{H.-F.}},
\bauthor{\bsnm{Zhang}, \binits{Q.}},
\bauthor{\bsnm{Pan}, \binits{J.-W.}}:
\batitle{Free-space dissemination of time and frequency with $10^{-19}$ instability over 113~km}.
\bjtitle{Nature}
\bvolume{610},
\bfpage{661}--\blpage{666}
(\byear{2022})
\doiurl{10.1038/s41586-022-05228-5}
\end{barticle}
\endbibitem

\bibitem[\protect\citeauthoryear{Gozzard et~al.}{2022}]{goz22}
\begin{barticle}
\bauthor{\bsnm{Gozzard}, \binits{D.R.}},
\bauthor{\bsnm{Howard}, \binits{L.A.}},
\bauthor{\bsnm{Dix-Matthews}, \binits{B.P.}},
\bauthor{\bsnm{Karpathakis}, \binits{S.F.E.}},
\bauthor{\bsnm{Gravestock}, \binits{C.T.}},
\bauthor{\bsnm{Schediwy}, \binits{S.W.}}:
\batitle{Ultrastable free-space laser links for a global network of optical atomic clocks}.
\bjtitle{Phys. Rev. Lett.}
\bvolume{128},
\bfpage{020801}
(\byear{2022})
\doiurl{10.1103/PhysRevLett.128.020801}
\end{barticle}
\endbibitem

\bibitem[\protect\citeauthoryear{Mehlst{\"a}ubler et~al.}{2018}]{meh18}
\begin{barticle}
\bauthor{\bsnm{Mehlst{\"a}ubler}, \binits{T.}},
\bauthor{\bsnm{Grosche}, \binits{G.}},
\bauthor{\bsnm{Lisdat}, \binits{C.}},
\bauthor{\bsnm{Schmidt}, \binits{P.}},
\bauthor{\bsnm{Denker}, \binits{H.}}:
\batitle{Atomic clocks for geodesy}.
\bjtitle{Rep. Prog. Phys.}
\bvolume{81},
\bfpage{064401}
(\byear{2018})
\doiurl{10.1088/1361-6633/aab409}
\end{barticle}
\endbibitem

\bibitem[\protect\citeauthoryear{{{I}nternational {C}lock and {O}scillator {N}etworking {C}ollaboration}}{2024}]{icon24}
\begin{botherref}
\oauthor{\bsnm{{{I}nternational {C}lock and {O}scillator {N}etworking {C}ollaboration}}}:
International comparison of optical frequencies with transportable optical lattice clocks
(2024).
\doiurl{10.48550/arXiv.2410.22973}
\end{botherref}
\endbibitem

\bibitem[\protect\citeauthoryear{Denker et~al.}{2019}]{den19}
\begin{botherref}
\oauthor{\bsnm{Denker}, \binits{H.}},
\oauthor{\bsnm{Timmen}, \binits{L.}},
\oauthor{\bsnm{V\"olksen}, \binits{C.}}:
Report on levelling and {GNSS} results for stations on the {MPQ} campus in {Garching}.
Technical report
(2019).
\doiurl{10.15488/9163}
\end{botherref}
\endbibitem

\bibitem[\protect\citeauthoryear{M{\"a}kinen}{2021}]{mae21}
\begin{barticle}
\bauthor{\bsnm{M{\"a}kinen}, \binits{J.}}:
\batitle{The permanent tide and the {International} {Height} {Reference} {Frame} {IHRF}}.
\bjtitle{J. Geod.}
\bvolume{95}(\bissue{9}),
\bfpage{106}
(\byear{2021})
\doiurl{10.1007/s00190-021-01541-5}
\end{barticle}
\endbibitem

\bibitem[\protect\citeauthoryear{Ihde et~al.}{2008}]{ihd08}
\begin{botherref}
\oauthor{\bsnm{Ihde}, \binits{J.}},
\oauthor{\bsnm{M{\"a}kinen}, \binits{J.}},
\oauthor{\bsnm{Sacher}, \binits{M.}}:
Conventions for the {D}efinition and {R}ealization of a {E}uropean {V}ertical {R}eference {S}ystem {(EVRS)}--{EVRS C}onventions 2007.
EVRS Conv,
1--10
(2008)
\end{botherref}
\endbibitem

\bibitem[\protect\citeauthoryear{Voigt et~al.}{2016}]{voi16}
\begin{barticle}
\bauthor{\bsnm{Voigt}, \binits{C.}},
\bauthor{\bsnm{Denker}, \binits{H.}},
\bauthor{\bsnm{Timmen}, \binits{L.}}:
\batitle{Time-variable gravity potential components for optical clock comparisons and the definition of international time scales}.
\bjtitle{Metrologia}
\bvolume{53}(\bissue{6}),
\bfpage{1365}
(\byear{2016})
\doiurl{10.1088/0026-1394/53/6/1365}
\end{barticle}
\endbibitem

\bibitem[\protect\citeauthoryear{Torge}{1975}]{tor75}
\begin{bbook}
\bauthor{\bsnm{Torge}, \binits{W.}}:
\bbtitle{Geod{\"a}sie}.
\bpublisher{Walter de Gruyter},
\blocation{Berlin}
(\byear{1975}).
\doiurl{10.1515/9783110198171}
\end{bbook}
\endbibitem

\bibitem[\protect\citeauthoryear{Moritz}{1980}]{mor80}
\begin{barticle}
\bauthor{\bsnm{Moritz}, \binits{H.}}:
\batitle{Geodetic reference system 1980}.
\bjtitle{Bull. G{\'e}od{\'e}sique}
\bvolume{54}(\bissue{3}),
\bfpage{395}--\blpage{405}
(\byear{1980})
\doiurl{10.1007/BF02521480}
\end{barticle}
\endbibitem

\bibitem[\protect\citeauthoryear{{IERS Earth Orientation Centre}}{2026}]{IERS_EOP20C04_1962now}
\begin{botherref}
\oauthor{\bsnm{{IERS Earth Orientation Centre}}}:
{IERS EOP 20u24 C04 (IAU2000A, dX, dY) 0hUTC Combined Series (1962--now)}.
Observatoire de Paris.
Data file \texttt{eopc04.1962-now}, accessed 2026-08-31
(2026).
\url{https://obspm.fr}
\end{botherref}
\endbibitem

\bibitem[\protect\citeauthoryear{Herbers et~al.}{2022}]{her22}
\begin{barticle}
\bauthor{\bsnm{Herbers}, \binits{S.}},
\bauthor{\bsnm{H\"{a}fner}, \binits{S.}},
\bauthor{\bsnm{D\"{o}rscher}, \binits{S.}},
\bauthor{\bsnm{L\"{u}cke}, \binits{T.}},
\bauthor{\bsnm{Sterr}, \binits{U.}},
\bauthor{\bsnm{Lisdat}, \binits{C.}}:
\batitle{Transportable clock laser system with an instability of $1.6 \times 10^{-16}$}.
\bjtitle{Opt. Lett.}
\bvolume{47}(\bissue{20}),
\bfpage{5441}--\blpage{5444}
(\byear{2022})
\doiurl{10.1364/OL.470984}
\end{barticle}
\endbibitem

\bibitem[\protect\citeauthoryear{Matei et~al.}{2017}]{mat17a}
\begin{barticle}
\bauthor{\bsnm{Matei}, \binits{D.G.}},
\bauthor{\bsnm{Legero}, \binits{T.}},
\bauthor{\bsnm{H\"afner}, \binits{S.}},
\bauthor{\bsnm{Grebing}, \binits{C.}},
\bauthor{\bsnm{Weyrich}, \binits{R.}},
\bauthor{\bsnm{Zhang}, \binits{W.}},
\bauthor{\bsnm{Sonderhouse}, \binits{L.}},
\bauthor{\bsnm{Robinson}, \binits{J.M.}},
\bauthor{\bsnm{Ye}, \binits{J.}},
\bauthor{\bsnm{Riehle}, \binits{F.}},
\bauthor{\bsnm{Sterr}, \binits{U.}}:
\batitle{$1.5~\mu$m lasers with sub-{10 mHz} linewidth}.
\bjtitle{Phys. Rev. Lett.}
\bvolume{118},
\bfpage{263202}
(\byear{2017})
\doiurl{10.1103/PhysRevLett.118.263202}
\end{barticle}
\endbibitem

\bibitem[\protect\citeauthoryear{Predehl et~al.}{2012}]{pre12}
\begin{barticle}
\bauthor{\bsnm{Predehl}, \binits{K.}},
\bauthor{\bsnm{Grosche}, \binits{G.}},
\bauthor{\bsnm{Raupach}, \binits{S.M.F.}},
\bauthor{\bsnm{Droste}, \binits{S.}},
\bauthor{\bsnm{Terra}, \binits{O.}},
\bauthor{\bsnm{Alnis}, \binits{J.}},
\bauthor{\bsnm{Legero}, \binits{T.}},
\bauthor{\bsnm{H\"ansch}, \binits{T.W.}},
\bauthor{\bsnm{Udem}, \binits{T.}},
\bauthor{\bsnm{Holzwarth}, \binits{R.}},
\bauthor{\bsnm{Schnatz}, \binits{H.}}:
\batitle{A 920-kilometer optical fiber link for frequency metrology at the 19th decimal place}.
\bjtitle{Science}
\bvolume{336}(\bissue{6080}),
\bfpage{441}--\blpage{444}
(\byear{2012})
\doiurl{10.1126/science.1218442}
\end{barticle}
\endbibitem

\bibitem[\protect\citeauthoryear{Raupach et~al.}{2015}]{rau15}
\begin{barticle}
\bauthor{\bsnm{Raupach}, \binits{S.M.F.}},
\bauthor{\bsnm{Koczwara}, \binits{A.}},
\bauthor{\bsnm{Grosche}, \binits{G.}}:
\batitle{Brillouin amplification supports $1\times10^{-20}$ accuracy in optical frequency transfer over 1400~km of underground fibre}.
\bjtitle{Phys. Rev. A}
\bvolume{92},
\bfpage{021801}
(\byear{2015})
\doiurl{10.1103/PhysRevA.92.021801}
\end{barticle}
\endbibitem

\bibitem[\protect\citeauthoryear{Ger\v{s}l et~al.}{2015}]{ger15a}
\begin{barticle}
\bauthor{\bsnm{Ger\v{s}l}, \binits{J.}},
\bauthor{\bsnm{Delva}, \binits{P.}},
\bauthor{\bsnm{Wolf}, \binits{P.}}:
\batitle{Relativistic corrections for time and frequency transfer in optical fibres}.
\bjtitle{Metrologia}
\bvolume{52}(\bissue{4}),
\bfpage{552}
(\byear{2015})
\doiurl{10.1088/0026-1394/52/4/552}
\end{barticle}
\endbibitem

\bibitem[\protect\citeauthoryear{Benkler et~al.}{2019}]{ben19}
\begin{barticle}
\bauthor{\bsnm{Benkler}, \binits{E.}},
\bauthor{\bsnm{Lipphardt}, \binits{B.}},
\bauthor{\bsnm{Puppe}, \binits{T.}},
\bauthor{\bsnm{Wilk}, \binits{R.}},
\bauthor{\bsnm{Rohde}, \binits{F.}},
\bauthor{\bsnm{Sterr}, \binits{U.}}:
\batitle{End-to-end topology for fiber comb based optical frequency transfer at the $10^{-21}$ level}.
\bjtitle{Opt. Express}
\bvolume{27}(\bissue{25}),
\bfpage{36886}--\blpage{36902}
(\byear{2019})
\doiurl{10.1364/OE.27.036886} .
\bcomment{also see erratum \cite{ben20}}
\end{barticle}
\endbibitem

\bibitem[\protect\citeauthoryear{Lodewyck et~al.}{2020}]{lod20}
\begin{barticle}
\bauthor{\bsnm{Lodewyck}, \binits{J.}},
\bauthor{\bsnm{Le~Targat}, \binits{R.}},
\bauthor{\bsnm{Pottie}, \binits{P.-E.}},
\bauthor{\bsnm{Benkler}, \binits{E.}},
\bauthor{\bsnm{Koke}, \binits{S.}},
\bauthor{\bsnm{Kronj\"ager}, \binits{J.}}:
\batitle{Universal formalism for data sharing and processing in clock comparison networks}.
\bjtitle{Phys. Rev. Research}
\bvolume{2},
\bfpage{043269}
(\byear{2020})
\doiurl{10.1103/PhysRevResearch.2.043269}
\end{barticle}
\endbibitem

\bibitem[\protect\citeauthoryear{}{2008}]{gum08}
\begin{botherref}
Evaluation of Measurement Data -- Guide to the Expression of Uncertainty in Measurement.
JCGM 100:2008
(2008).
\url{http://www.bipm.org/utils/common/documents/jcgm/JCGM_100_2008_E.pdf}
\end{botherref}
\endbibitem

\bibitem[\protect\citeauthoryear{Benkler et~al.}{2020}]{ben20}
\begin{barticle}
\bauthor{\bsnm{Benkler}, \binits{E.}},
\bauthor{\bsnm{Lipphardt}, \binits{B.}},
\bauthor{\bsnm{Puppe}, \binits{T.}},
\bauthor{\bsnm{Wilk}, \binits{R.}},
\bauthor{\bsnm{Rohde}, \binits{F.}},
\bauthor{\bsnm{Sterr}, \binits{U.}}:
\batitle{End-to-end topology for fiber comb based optical frequency transfer at the $10^{-21}$ level: erratum}.
\bjtitle{Opt. Express}
\bvolume{28}(\bissue{10}),
\bfpage{15023}--\blpage{15024}
(\byear{2020})
\doiurl{10.1364/OE.27.036886}
\end{barticle}
\endbibitem

\end{thebibliography}

%
\end{document}